\documentclass[fleqn,usenatbib]{mnras}

\usepackage[T1]{fontenc}
\usepackage{ae,aecompl}
\PassOptionsToPackage{pdfpagelabels=true}{hyperref}

\usepackage{graphicx}	
\usepackage{amsmath}	
\usepackage{amssymb}	
\usepackage{nccmath}

\usepackage{mathptmx}
\usepackage{mathrsfs}
\usepackage{epsfig}
\usepackage{pdflscape}
\usepackage{pifont}
\usepackage{afterpage}
\usepackage{paralist}
\usepackage[caption=false]{subfig}
\usepackage[flushleft]{threeparttable}
\usepackage{float}
\usepackage{array}
\newcolumntype{P}[1]{>{\centering\arraybackslash}p{#1}}
\usepackage{lscape}
\usepackage[section]{placeins}
\usepackage{url}
\usepackage{natbib}
\usepackage{gensymb}
\usepackage{figure}
\usepackage{comment}
\usepackage{multirow}

\defcitealias{harwood13}{H13}
\defcitealias{harwood15}{H15}
\defcitealias{harw16}{H16}
\defcitealias{harw17}{H17}

\title[The spectral age problem]{Investigating the spectral age problem with powerful radio galaxies}

\author[V.H. Mahatma et al.]{Vijay H. Mahatma,$^{1}$
Martin J. Hardcastle,$^{1}$
Judith H. Croston,$^{2}$
Jeremy Harwood,$^{1}$\newauthor
Judith Ineson$^{3}$
and Javier Moldon$^{4,5}$ 
\\
$^{1}$Centre for Astrophysics Research, School of Physics, Astronomy
\& Mathematics, University of Hertfordshire, College Lane, Hatfield
AL10 9AB, UK\\
$^{2}$School of Physical Sciences, The Open University, Walton Hall, Milton Keynes MK7 6AA, UK\\
$^{3}$School of Physics and Astronomy, University of Southampton, Southampton S17 1BJ, UK\\
$^{4}$Jodrell Bank Centre for Astrophysics, School of Physics and Astronomy,
University of Manchester, Manchester M13 9PL, UK
\\
$^{5}$Instituto de Astrof\'isica de Andaluc\'ia (IAA, CSIC), Glorieta de las Astronom\'ia, s/n, E-18008 Granada, Spain}

\date{Accepted 2019 December 02. Received 2019 December 02; in original form 2019 October 04}

\pubyear{2019}

\begin{document}
\label{firstpage}
\pagerange{\pageref{firstpage}--\pageref{lastpage}}
\maketitle

\begin{abstract}
The `spectral age problem' is our systematic inability to reconcile
the maximum cooling time of radiating electrons in the lobes of a
radio galaxy with its age as modelled by the dynamical evolution of
the lobes. While there are known uncertainties in the models that
produce both age estimates, `spectral' ages are commonly
underestimated relative to dynamical ages, consequently leading to
unreliable estimates of the time-averaged kinetic feedback of a
powerful radio galaxy. In this work we attempt to solve the spectral
age problem by observing two cluster-centre powerful radio galaxies;
3C320 and 3C444. With high-resolution broad-band Karl G. Jansky Very
Large Array observations of the radio sources and deep
\textit{XMM-Newton} and \textit{Chandra} observations of their hot
intra-cluster media, coupled with the use of an analytic model, we
robustly determine their spectral and dynamical ages. After finding
self-consistent dynamical models that agree with our observational
constraints, and accounting for sub-equipartition magnetic fields, we
find that our spectral ages are still underestimated by a factor of
two at least. Equipartition magnetic fields will underestimate the
spectral age by factors of up to $\sim$ 20. The turbulent mixing of
electron populations in the radio lobes is likely to be the main
remaining factor in the spectral age/dynamical age discrepancy, and must be accounted for in the study of large samples of powerful radio galaxies. 
\end{abstract}




\section{Introduction}
\subsection{Radio galaxies and their environments}
Radio galaxies, or radio-loud AGN (RLAGN), drive powerful jets of plasma into the environment of their host galaxies, inflating large radio-emitting lobes seen in radio observations. It is now well known that such large-scale radio emission (physical linear scales of a few Mpc for the largest sources, e.g. \citealt{dabh17}) originates from the process of accretion onto the supermassive black hole (SMBH) of the host galaxy, where the jets are launched. For the most powerful sources, which tend to be hosted by massive elliptical galaxies \citep{best04,best12,saba19}, the jets may escape their central dense environments and terminate in the inter-galactic (IGM) or intra-cluster medium (ICM). 

The jets of RLAGN can have profound implications for cosmic evolution. The widely postulated AGN feedback mechanism \citep[e.g.][]{crot06,wyle16} attempts to account for the shape of the galaxy mass function, which shows a steep decline of galaxies with increasing stellar mass. One of the mechanisms by which this can physically be explained is interstellar gas heating by the jets, which is suggested to be linked to the suppression of star formation in the most massive galaxies. While as yet there is no direct observational evidence for the population of RLAGN in general affecting their host-galaxy star-formation rates, there \textit{is} robust evidence for significant heating of the interstellar or intracluster gas by powerful jets as they drive shocks into their surrounding medium \citep[e.g. ][]{kraf03,cros07,cros09,liu19}. To understand what role this heating plays in the evolution of galaxies or clusters, it is important to understand and quantify the energetics of AGN during a radio-loud jet outburst, and how long they spend in this phase.

Of particular importance is the power output of a radio galaxy during
each outburst (the jet power), which is a useful quantity for models
of feedback processes \citep{hein07,gasp12_feedback}. Recently,
\cite{turn15} and \cite{hard19} inferred the bulk jet powers of the
population of observed RLAGN from different radio surveys, assuming a
distribution of radio galaxy environments based on cluster atmospheres
and a particular lifetime function. While such inferences may be made for
a population of sources, the jet power for any given source may
be determined observationally by directly probing the
  jet energetics. \cite{godf13} estimate jet powers using hotspot
  luminosities, while \cite{loba98} use the spatial shifts of the
  radio core in VLBI images, probing the parsec scale jets. To measure
  jet powers observationally for large samples, the
    simplest method in principle is to measure the total instantaneous energy output by the source ($E_{\text{total}}$), its lifetime ($t_{\text{age}}$) and taking their ratio ($Q_{\text{jet}}=E_{\text{total}}$ / $t_{\text{age}}$). Methods of estimating $E_{\text{tot}}$ and $t_{\text{age}}$ are discussed below.
\subsection{Energy estimates}
Methods of directly estimating $E_{\text{total}}$
have traditionally varied. For RLAGN in rich cluster environments, the
lobes may excavate cavities in the hot X-ray emitting ICM, leading to
large-scale decrements in the X-ray surface brightness, which enable
$pdV$ estimates of the work done by the lobes on their surroundings
\citep{raff06,birz04}. Such a calculation, however, may result in an
underestimate of the total energy output by the jets -- numerical
simulations by \cite{hakr13} and \cite{engl16} show that the energy
contained in the lobes (through radiating and non-radiating particles)
is estimated to be only half of the total energy ever injected by the
jets, where the remaining energy is transferred into shock-heating the
surrounding medium (see \citealt{osul11} for a list of uncertainties
regarding the use of cavity power estimates). Cavity power estimates
have been used to derive a linear scaling relationship between jet
power and radio luminosity, as found by \cite{cava10}, but such a
relation is bound to be unreliable for inferring the jet power for a
given radio luminosity, as physical parameters such as the environmental density and the source age must also control the observed radio luminosity. Multi-wavelength information describing the energetics of the radio lobes \textit{and} its surrounding shocked material, applicable to all radio galaxies, can give more reliable estimates of $E_{\text{total}}$. \cite{ines17} indeed use $E_{\text{total}}=2E_i$ where $E_i$ is the internal energy density of the lobes calculated directly using radio synchrotron and X-ray inverse-Compton measurements of a large sample of FR-II sources. Given a correct estimate of the source age, this would provide a reliable calculation of jet power.
\subsection{Age estimates}
Radio galaxy lifetime estimates can also be found in various ways: dynamical models which describe the speed of the radio jets, the lobes or their shocks have often been invoked in combination with the instantaneous source size to determine `dynamical ages' \citep{kais97,kais00}. As an alternative to dynamical ages, `spectral ageing' can be used to determine the radiative ages of the relativistic electron population in radio galaxies, and has been used extensively as a tool to derive source ages from observations \citep[e.g.][]{hargrave74,alex87,cari91,shul12} as well as from synthetic lobe radio spectra \citep{turn+18,turn18}. We discuss the usage and the validity of both methods below.
\subsubsection{Dynamical ages}
The determination of lobe advance speeds from observations has been
attempted with X-ray observations: powerful jets drive a shock into their
surrounding medium, heating their group or cluster media \citep{hein98,kraf07,cros09},
and so X-ray measurements of physical conditions allow estimates of shock speed \citep[e.g.][]{ines17}.
Alternatively, cavity power estimates often use the assumption that
the lobes have expanded at the sound speed
\citep{birz04,birz08,cava10}. Both these methods have a clear
drawback: the rate of expansion of the lobes is not constant, as radio
lobes in general decelerate with time due to the interaction with
their surrounding environment. Because the speed in the past is likely
to be higher than its instantaneous value, these models overestimate the dynamical age.

Analytic models or numerical simulations \citep[e.g.][]{hakr13,
  turn15,engl16,hard18} that aim to reproduce the observed properties
of a particular source can circumvent problems with instantaneous
measurements. However, numerical modelling requires jet power and
environmental information as an initial condition. The use of a suite
of different simulations to estimate jet power from observables would
be possible in principle but computationally very expensive.
\subsubsection{Spectral ages}
The radiative ageing of electrons in the lobes of powerful radio
galaxies occurs as follows. As the jet propagates into its
environment, a strong shock (often termed reverse shock) develops at
its termination point, which in turn accelerates particles to very
high energies (Lorentz factors of $\sim 10^{5}$ or higher)
through magnetic inhomogeneities in front of and behind the shock
\citep{bell78}. As a simple model, a single population of accelerated
electrons at the same energy will leave the acceleration region as the
jet advances, travelling downstream. These electrons, filling the
radio lobe, will then suffer energy losses via the synchrotron,
inverse-Compton and adiabatic expansion processes. The loss timescale for the synchrotron
process goes as $\tau \sim 1$/$E$ and hence higher energy electrons
radiate away their energy $E$ more rapidly than lower energy
electrons. This creates a characteristic observed radio spectrum which
is a power law at low frequencies and curves downwards at higher frequencies. The spectral age is given by 
\begin{ceqn}
\begin{equation}
t_{spec}=50.3\frac{B^{1/2}}{B^2+B^2_{CMB}}\left(\left(1+z\right)\nu_{b}\right)^{-1/2} \text{Myr}
\label{equation:specage}
\end{equation} 
\end{ceqn}
\citep{leah91}, where $B$ (nT) is the magnetic field strength in the lobes, $B_{\text{CMB}}$ (nT) is the magnetic field strength in the cosmic microwave background, $z$ is the sources' redshift and $\nu_b$ (GHz) is the `break' frequency where radiative losses begin to significantly affect the observed spectrum. Both the magnetic field strength and the break frequency are required parameters; the latter can be obtained directly from the observed radio spectrum.  

The reliability of this method to accurately determine the age of
radio galaxies is unclear. The spectral age is strongly dependent on
the lobe magnetic field, the value of which has often been estimated
using the argument that the energy density in the magnetic field is in
equipartition with that in the radiating particles. While this
assumption has been invoked for many years, inverse-Compton
measurements. at least for Fanaroff-Riley type II
\citep[FR-II;][]{FR74} sources, constrain the lobe magnetic field
strengths to be substantially lower than those implied by
equipartition \citep{cros05,ines17}. Furthermore, low-sensitivity,
narrow-bandwidth observations from the previous generation of radio
telescopes may not accurately trace the oldest radiating particles
since last injection -- a feature of the rapidly fading nature of
synchrotron radiation that also prevents detections of switched-off
radio galaxies even in sensitive low frequency surveys
\citep{brie16,brie17,maha18}.
\label{sec:equipartition}
\subsubsection{Outstanding questions} \label{sect:dynamical_ages}
Even with current generation radio telescopes such as the Karl G.
Jansky Very Large Array (VLA), with its high sensitivity and
broad-band capabilities, and our improved understanding of the
dynamical evolution and the energetics of the lobes, a discrepancy
between the spectral and dynamical age of sources is still observe
\citep{eile96,harwood13,harwood15}. Spectral ages are still being
underestimated with respect to dynamical ages. This implies that the
use of spectral ages to derive the jet power of radio-loud AGN would
overestimate the true kinetic power of the source, if the modelled
dynamical ages are correct. \cite{harwood13} and \cite{harwood15}
(hereafter \citetalias{harwood13} and \citetalias{harwood15},
respectively) investigated the spectral age/dynamical age problem of
bright FR-II radio galaxies, developing the {\sc brats} software package --
a tool that derives spectral ages from radio maps from spatial regions
determined on a pixel-by-pixel basis. This allows age estimates to be
obtained as a function of distance along the lobes on a truly resolved
basis, rather than integrating over large spatial regions as has been
done in the past \citep[e.g.][]{al-le87}. \citetalias{harwood15},
using the {\sc brats} package, still found that that their spectral ages can
be underestimated by an order of magnitude with respect to their
dynamical ages. Their use of equipartition estimates may have been the
primary driver in the discrepancy, but they also used a simple
power-law model for the environments; ideally environments would be
constrained by observations. Moreoever, subsequent studies have
suggested that even the faintest radio emission from the lobes may
trace a mixture of electron populations with different ages.
\cite{harw16,harw17} predicted that electron mixing would be a
contributing factor in determining spectral ages that may be biased
towards younger populations. \cite{turn+18} later confirmed these
predictions with numerical simulations, showing that such electron
mixing occurs progressively for the oldest material, strongly
affecting the observed spectrum, and hence the spectral age. It is
important to quantify observationally the effect of electron mixing.

A further uncertainty in spectral ageing comes from the choice of
injection index, $\alpha_{inj}$, the index of the initial power law
injected by particle acceleration (\citetalias{harwood13},
\citetalias{harwood15}). Observations of hotspots have given values of
around 0.5 to 0.6 for $\alpha_{inj}$ \citep{meisen89,carilli91} while
shock theory predicts a value of 0.5 \citep{long11}. However,
\citetalias{harwood13} and \citetalias{harwood15} found
$\alpha_{inj}\sim 0.8$, while \cite{harw17} find $\alpha_{inj}\sim
0.7$ with broader coverage at low frequencies, suggesting that
absorption processes or additional acceleration mechanisms may affect
fitted measurements. While a lack of frequency coverage at low
frequencies ($\sim 1.5$ GHz) may cause the models to predict steeper
injection indices, it may simply be the case that FR-II radio galaxies
have a large spread of $\alpha_{inj}$. Further studies are needed to
understand the value and distribution of $\alpha_{inj}$.

In general, it is crucial to understand whether physical or observational effects drive the dynamical/spectral age discrepancy. Deep observations of cluster-centre radio galaxies displaying large-scale shocks being driven by the expanding radio lobes into the surrounding environment, as well as high resolution and sensitive broad-bandwidth radio maps of the radio sources are necessary. This will allow us to capture the faintest and most aged electron populations in the radio lobes for a spectral ageing study, and with a reliable model for the dynamical advance of the radio lobes, we may investigate the cause of the spectral age/dynamical age discrepancy in detail.

There are, however, few observations of powerful sources driving
shocks in their surrounding medium, due to the rarity of powerful sources at low redshift. So far, in addition to Cygnus A, there are only a handful of FR-II sources with X-ray observations showing a central shock being driven by the lobes, while a much larger population of the lower-power FR-I systems interacting with the central parts of the cluster are known. Recently, VLA observations of the nearby young FR-I radio galaxy NGC3801, which drives a shock seen in the X-ray \citep{hees14}, have shown that the spectral and dynamical ages agree to within a factor of two. However, there is also significant evidence for in-situ particle acceleration and/or mixing of the electron population in the lobes. It is unclear how well this generalizes to much more powerful and physically larger objects such as FR-IIs which are better understood in terms of their dynamics -- classical doubles are routinely modelled, both analytically and numerically, to predict their physical properties as well as the impact onto their environments \citep[e.g.][]{sche74,kais97,hakr13,engl16}. As the predictions of these models are tested by detailed observations, the subsequent refinement of these models may also be used to predict properties of radio galaxies and their impact at high redshift. Low-redshift observations are ideal for these studies due to the sensitivity and resolution requirements.

In this paper, we present high-resolution. sensitive radio and X-ray
observations of two FR-II cluster-centre radio galaxies in order to
investigate the spectral age/dynamical age problem. By constraining
the source physical parameters using these observations, we determine
spectral ages using the techniques of \citetalias{harwood13} and
dynamical ages using a analytic model to investigate the spectral
age/dynamical age problem. In Section \ref{sect:observations} we
describe our observations and their data reduction. Section
\ref{sect:analysis} describes the techniques we employed to extract
physical parameters of our sources, the fitting of spectral ages to
our observations and the analytic modelling leading to robust
dynamical ages. In Section \ref{sect:discussion} we summarise our
results and discuss their implications, followed by our main
conclusions in Section \ref{sect:conclusions}. Some of the results
presented in this paper have been reported previously as part of a
research degree thesis \citep{mscthesis}.

Throughout this paper we define the spectral index $\alpha$ in the sense $S\propto\nu^{-\alpha}$. We use a $\Lambda$CDM cosmology in which $H_0 = 71$ kms$^{-1}$Mpc$^{-1}$, $\Omega_m$ = 0.27 and $\Omega_{\Lambda}$ = 0.73.
\section{Observations} \label{sect:observations}
\begin{figure*}
    \centering
    \subfloat{\includegraphics[scale=0.53, trim={2cm 8.5cm 2cm 6.5cm}, clip]{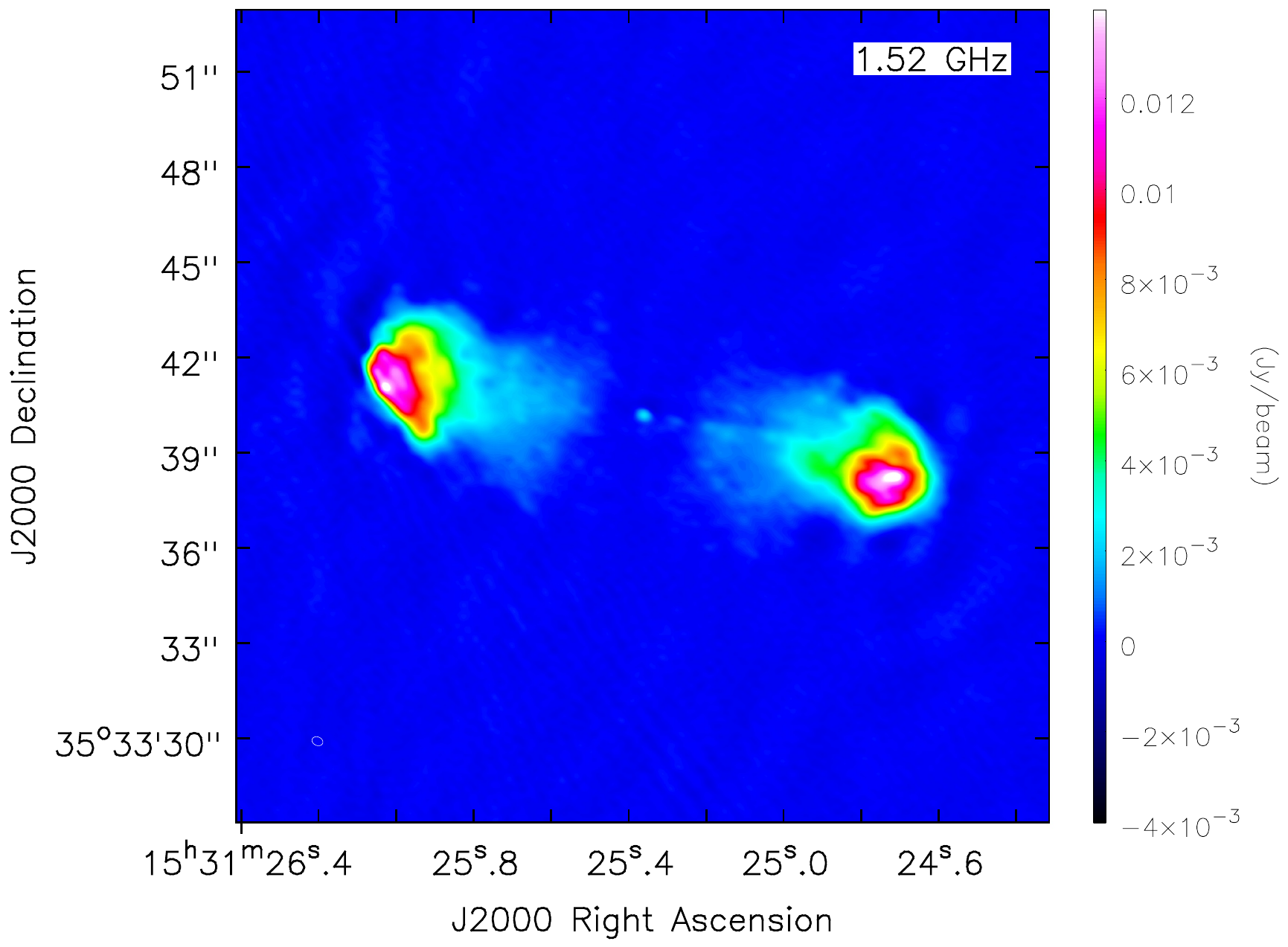}}
    \subfloat{\includegraphics[scale=0.53, trim={2cm 8.5cm 2cm 6.5cm}, clip]{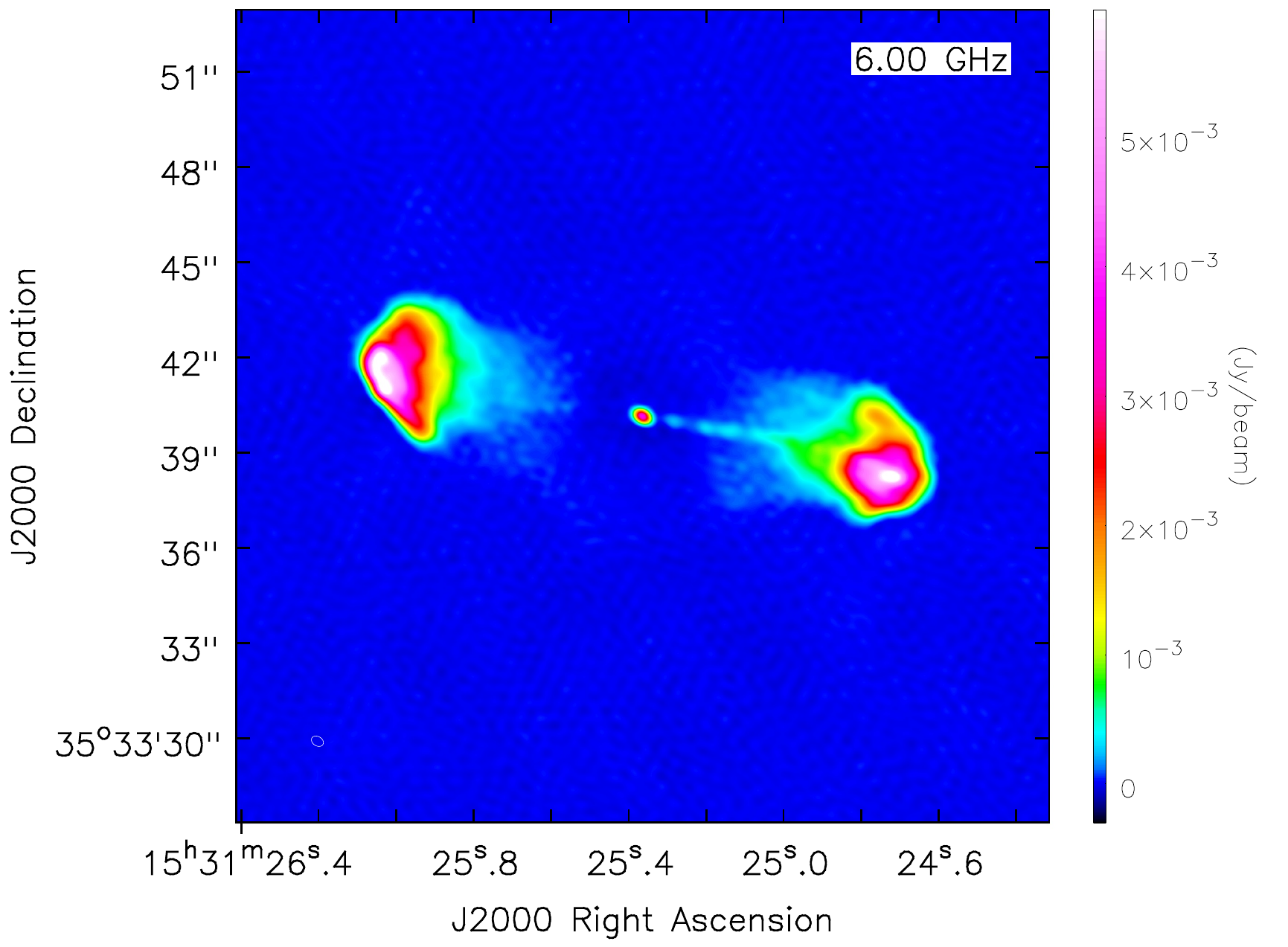}}\\
    \subfloat{\includegraphics[scale=0.53, trim={2cm 8.5cm 2cm 6.5cm}, clip]{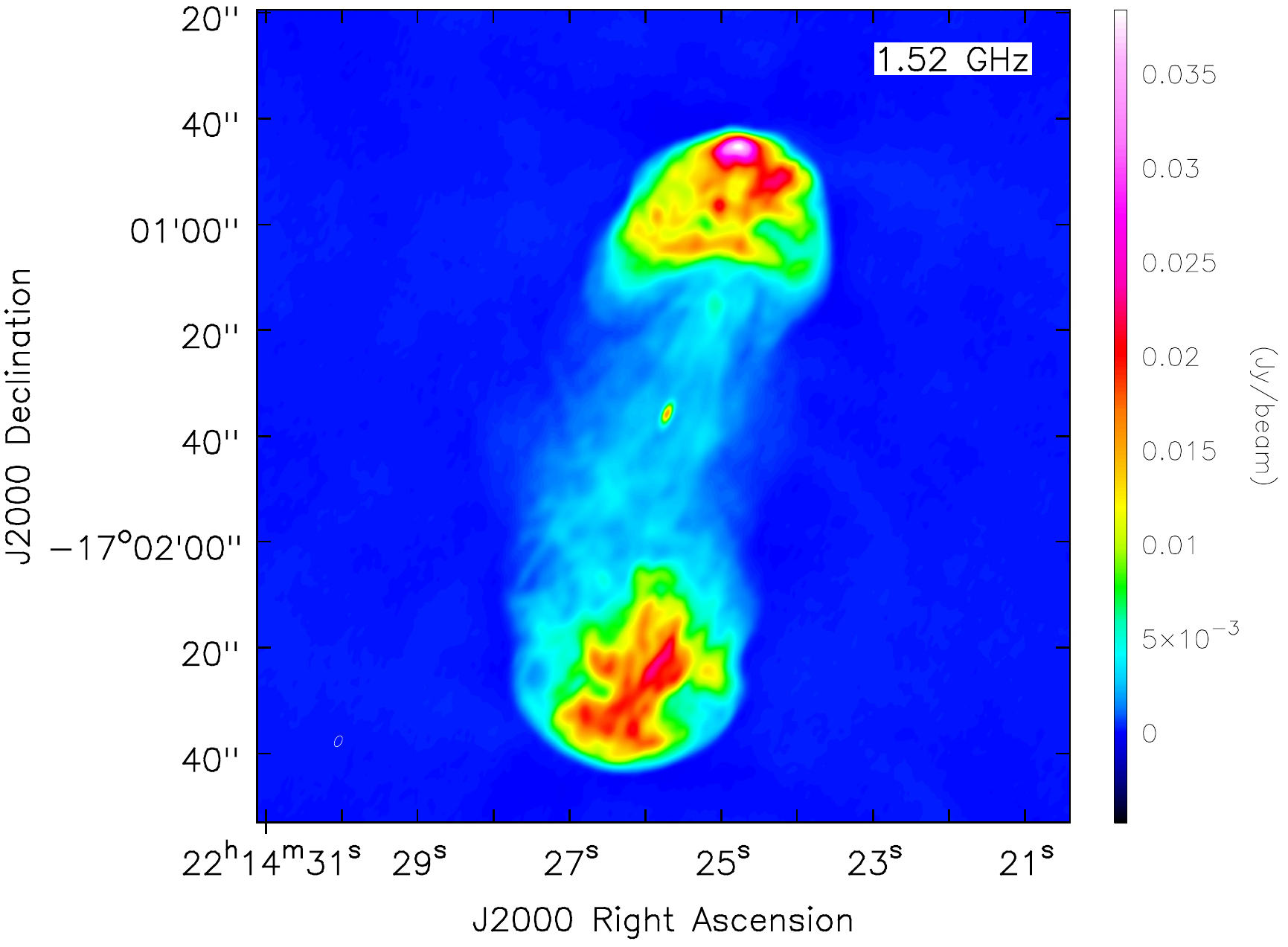}}
    \subfloat{\includegraphics[scale=0.53, trim={2cm 8.5cm 2cm 6.5cm}, clip]{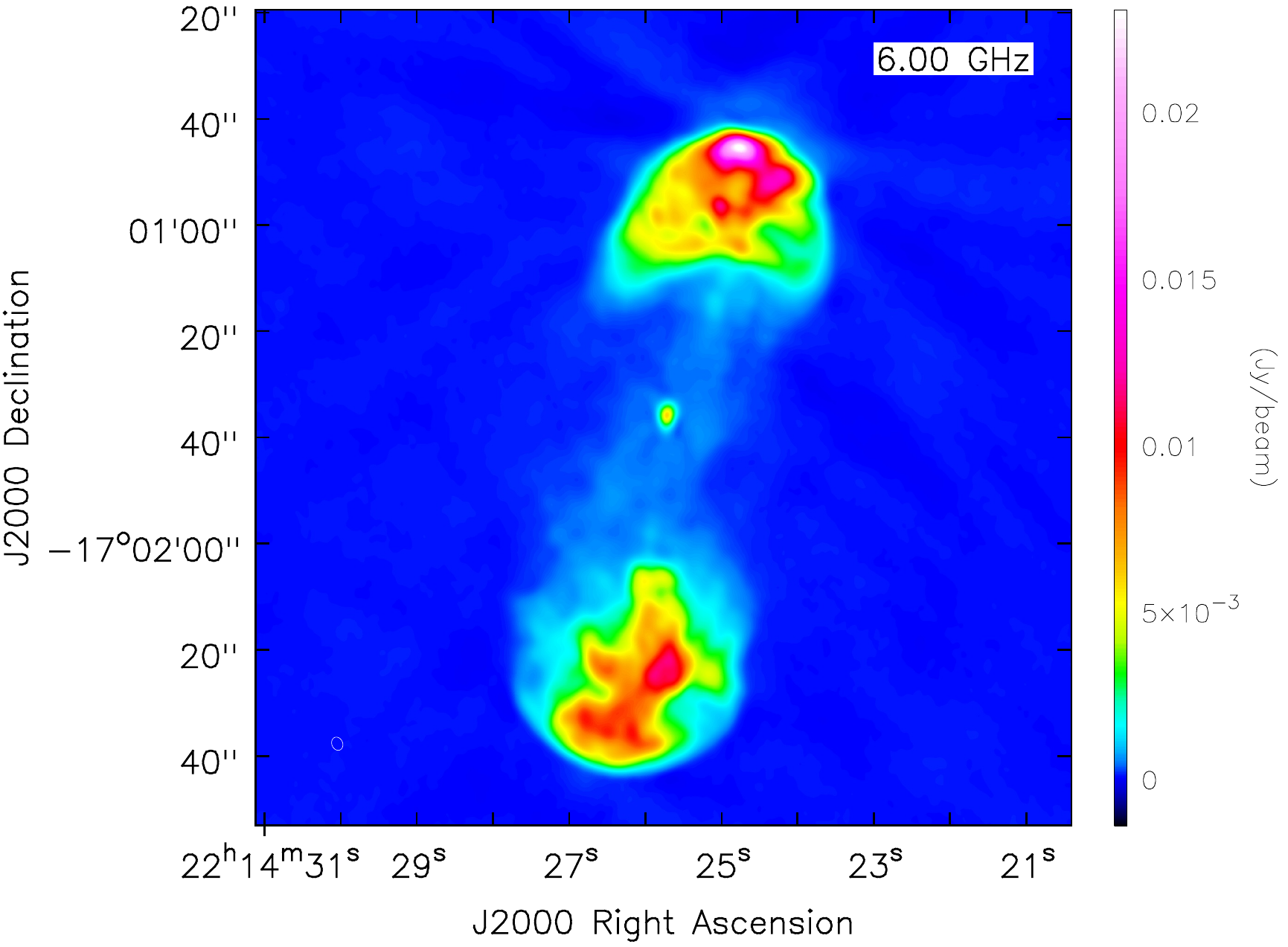}}
 \caption{VLA images of 3C320 (upper row) and 3C444 (lower row), at L-band/1.5 GHz (left column) and C-band/6 GHz (right column). Images shown are the final self-calibrated and array-combined broad-band data sets (see text). For 3C320 the beam sizes are $0.35\times0.27$ arcsec (left) and $0.57\times0.40$ arcsec (right). For 3C444 the beam sizes are $2.2\times1.3$ arcsec (left) and $2.6\times2.1$ arcsec (right). Beam shapes are shown on the lower left of each image panel. The pixel sizes are 0.1 arcsec for 3C320 (both), and 0.3 arsec (left) and 0.4 arcsec (right) for 3C444. Residual rms levels are$\sim 50$ $\mu$Jy beam$^{-1}$ (left) and $\sim 47$ $\mu$Jy beam$^{-1}$ (right) for 3C320, and $\sim 25$ $\mu$Jy beam$^{-1}$ (left) and $\sim 15$  $\mu$Jy beam$^{-1}$ (right) for 3C444. The colour bar in each image has scaled appropriately to display both compact and diffuse structures.}
\label{figure:radio_obs}
\end{figure*}

We require good quality observations of source environments in order to model their dynamical evolution. To add to the difficulty in obtaining such extensive observations, the fact that not all radio galaxies with the best radio observations necessarily drive observable shocks in their external medium means that the number of potential sources available to be observed is severely limited, particularly at low-redshift where FR-II sources are less common in dense environments. It is more efficient to initially cross-match bright sources found in radio surveys with X-ray surveys, followed by targeted deep X-ray observations of the shock, followed by high resolution and broad-band radio observations, rather than vice versa.
\begin{table*}
	\caption{X-ray observations of 3C320 and 3C444. The observation ID refers to IDs for the multiple data sets taken, if any.}
	\centering
    \begin{tabular}{P{1.2cm}P{1.7cm}P{2.8cm}P{1.2cm}P{2.5cm}P{1.8cm}P{1.5cm}}
    \hline
    Source & Telescope & Observation ID & Chip & Energy band \newline (keV) & Observation date & Duration \newline (ks) \\ \hline
    3C320 & \textit{Chandra} & 16130,16613 & ACIS-S & 0.5-7.0 & 2014-05-11 & 110 \\ \hline
    3C444 & \textit{XMM-Newton} & 0691840101 & MOS1, MOS2, PN & 0.3-7.0 & 2012-11-16 & 134 \\ 
    \hspace{1mm} & \textit{Chandra} & 15091 & ACIS-S & 0.5-7.0 & 2014-04-12 & 164 \\
    \hline 
    \end{tabular} \label{table:xrayobs}
\end{table*}
\begin{table*}
	\caption{Radio observation details of 3C320 and 3C444. The `Project ID' refers to the project name for the VLA and e-MERLIN data. `Array' gives the VLA array configuration used for each observation at L-band (1.5 GHz) and C-band (6 GHz), or the antennas used in the case of the e-MERLIN observations. The `Duration' refers to the approximate total observation time including all the calibrator sources.}
	\centering
    \begin{tabular}{P{1.2cm}P{2.8cm}P{1.2cm}P{2.5cm}P{2cm}P{1.5cm}P{2.5cm}}
    \hline
    Source & Project ID & Array & Frequency \newline (GHz) & Observation date & Duration \newline (hrs) & Flux Calibrator \\ \hline
    3C320 & 15A-420 (VLA) & \hspace{1mm} A \newline B & \hspace{3mm} 1.5, 6 \newline 6 & \ 01/08/15 \newline 15/02/15 & \hspace{2mm} 4 \newline 1.5 & 3C286 \\
    & CY4223 (e-MERLIN) & Mk2,Pi,Da,\newline Kn,De,Cm & 1.5 & 22/03/18 & 16 & 3C286 \\ \hline 
    3C444 & 15A-420 (VLA) & \hspace{1.3mm} A \newline \hspace*{1.5mm} B \newline C & \hspace{4mm} 1.5 \newline \hspace*{3mm} 1.5, 6 \newline C & \hspace{0.3mm} 18/06/15 \newline 07/02/15 \newline 31/01/16 & \hspace{2mm} 3.5 \newline \hspace*{1.5mm} 4 \newline 1.5 & 3C48 \\ \hline 
    \end{tabular} \label{table:radioobs}
\end{table*}
3C320 and 3C444 were both selected initially from unbiased \textit{Chandra} X-ray snapshot surveys of powerful radio galaxies by \cite{massaro13} and \cite{mingo14}, respectively, showing clear shock signatures (i.e surface brightness jumps around the lobes). Both sources were thus selected for deeper follow-up X-ray observations using \textit{Chandra} and \textit{XMM-Newton} for 3C320 and 3C444, respectively. For the radio follow-up, both targets were observed with the VLA in multiple array configurations for good $uv$ coverage and for broad-bandwidth data at 1.5 GHz (L-band; 1 GHz bandwidth) and at 6 GHz (C-band; 4 GHz bandwidth). Below, we describe the observations in detail for each source, followed by a description of the imaging processes.
\subsection{3C320}
3C320 is an intermediate redshift ($z=0.342$) classical FR-II radio
galaxy, hosted by a brightest cluster galaxy located in the centre of
a rich cluster as inferred from its X-ray emission \citep{massaro13}.
3C320 was selected as one of the brightest cluster galaxies (BCGs) in
its respective survey and followed up with deep \textit{Chandra}
observations, summarized in Table \ref{table:xrayobs}. These observations, taken for the present project, have already been analysed by \cite{vags19}, who find large X-ray cavities in the ICM associated with the radio lobes, and weak shocks surrounding the lobes.

The corresponding radio observations were made using the VLA at L-band in A-configuration (1.3 arcsec synthesized beam) and with e-MERLIN\footnote{\url{http://www.e-merlin.ac.uk}} for longer baseline data\footnote{Without the Lovell Telescope.} (0.15 arcsec synthesized beam), leading to a comparable combined angular resolution ($\sim 0.5$ arcsec) to that provided with the C-band observations, made with the combined VLA A-configuration and B-configuration data ($\sim 0.7$ arcsec beam). These radio observations are summarised in Table \ref{table:radioobs}.

\subsection{3C444}
\begin{figure*}
\centering
\subfloat{\includegraphics[scale=0.47, trim = {1.4cm 2cm 1.65cm 6.8cm}, clip]{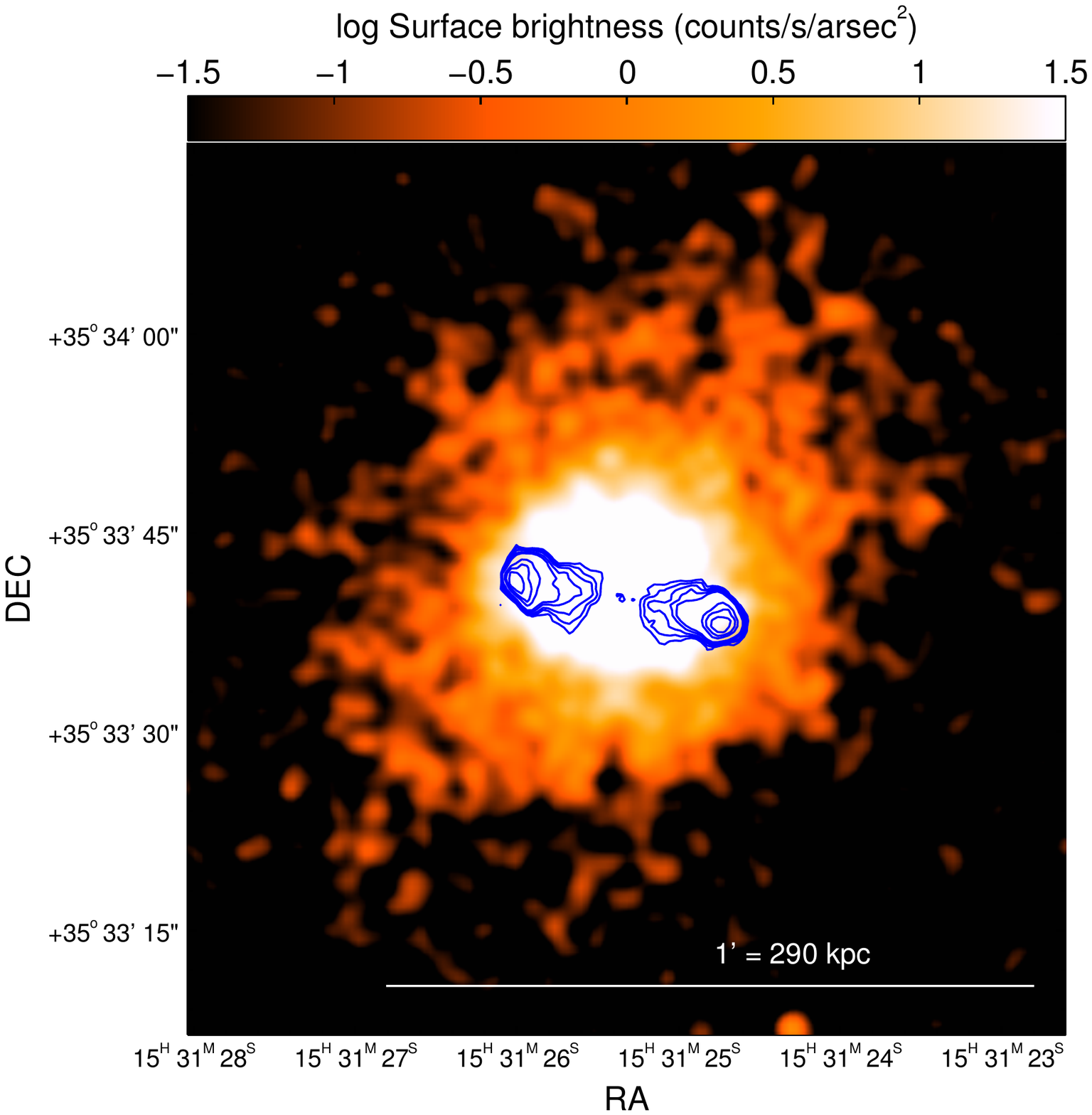}}
\subfloat{\includegraphics[scale=0.47, trim = {2.3cm 2cm 1.3cm 6.9cm}, clip]{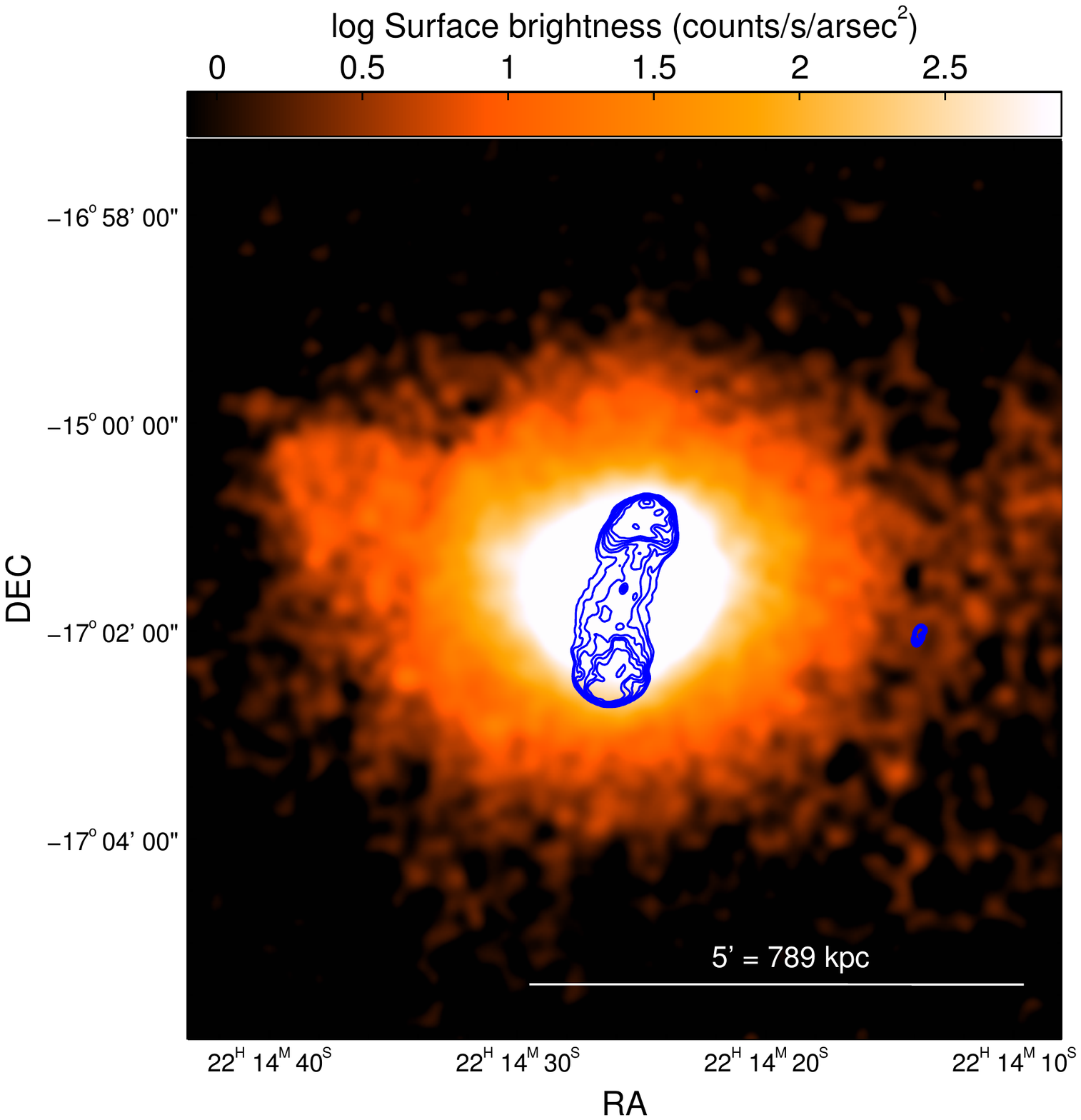}}
\caption{X-ray \textit{Chandra} (left) and \textit{XMM-Newton} (right) observations tracing the hot gas of the ICM surrounding 3C320 and 3C444, respectively. Images have been scaled using a logarithmic transfer function, allowing the sharp X-ray surface brightness increase around the radio source (the shock) to be seen clearly. Overlaid are 1.5 GHz contours of the radio source given by our VLA observations. The contour levels in this figure have been chosen to emphasize the radio structures clearly.}
\label{figure:clusters}
\end{figure*}

3C444 is a nearby ($z=0.153$) FR-II radio galaxy, hosted by an elliptical galaxy surrounded by a diffuse halo \citep[as detected with Gemini GMOS-S;][]{ramo11} at the centre of the cluster Abell 3847. Direct evidence for shock heating by the lobes of 3C444 probed by X-ray observations of the hot ICM has been given by \cite{cros11} and \cite{vags17} -- in the former study, 3C444 was selected serendipitously as part of a programme observing the southern 2Jy sample of radio galaxies \citep{morg93}. 

Unlike the case for 3C320, the snapshot \textit{Chandra} data for 3C444 were deep enough to constrain the shock and \cite{cros11} obtained a dynamical age, modelling the instantaneous shock Mach number directly using the X-ray emission. While these data are sufficient to measure the central shock properties, as analysed in Section \ref{sect:shock_measurements}, for our analytic modelling (Section \ref{sect:modelling}) we required more sensitive observations displaying the larger-scale outer environment around the shock. Deeper observations were made with \textit{XMM-Newton} of the larger-scale structure of the ICM, and are used to model the predicted central ICM properties in Section \ref{sect:analysis}. The \textit{Chandra} and \textit{XMM-Newton} observations are summarized in Table \ref{table:xrayobs}.

For the follow-up radio observations, the longest VLA baselines were sufficient to resolve the small scale structures (i.e the radio core and hotspots) and we did not require the longer baselines of e-MERLIN, owing to the fact that 3C444 is larger than 3C320 in angular size. At L-band, 3C444 was observed in the A and B-configuration, and at C-band, we used the B and C-configuration. The observations are summarized in Table \ref{table:radioobs}.
\subsection{Data reduction and imaging}
\subsubsection{Radio data}
For the VLA data, the data reduction was performed using the \textit{Common Astronomy Software Applications} (CASA) package version 4.7.1, in the standard manner, for both 3C320 and 3C444. The measurement sets for each VLA array at each observing frequency (L and C band) were reduced individually and manually using the standard CASA routines. The 3C320 e-MERLIN L-band data set was reduced using the e-MERLIN CASA pipeline\footnote{\url{https://github.com/e-merlin/eMERLIN_CASA_pipeline}}. Prior to any data reduction tasks the flagging framework AOFlagger \citep{offr12} was used to flag channels with radio-frequency interference (RFI). Flagging bad antennas that were stated in the VLA observers log was carried out only if erratic phase and amplitude gain solutions were derived for the same antennas. 

Prior to initial phase and bandpass calibration, the most recent model flux densities of 3C286 and 3C48 (for 3C320 and 3C444, respectively) were applied to the flux calibrator observations using the scale of \cite{perl13}. Antenna delay, bandpass, phase and amplitude gain calibrations were performed, applying the previous calibration solutions on the fly for better calibration. The CASA \texttt{rflag} algorithm was also used on the data sets to remove residual RFI after applying the calibration tables.

The CASA task \texttt{CLEAN} \citep{hogbom} was used as the deconvolution algorithm for imaging. This task was used to combine the various VLA array data at L and C-band in the $uv$ plane, as well as the L-band VLA and e-MERLIN data for 3C320, during the deconvolution process to produce high-fidelity broad bandwidth images, as shown in Figure \ref{figure:radio_obs}. Most of the default CASA \texttt{CLEAN} parameters were kept, bearing in mind that we require emphasis on image fidelity, showing both compact structures and the diffuse large scale (ageing) structure. Hence, we performed multi-scale cleaning and multi-frequency synthesis, owing to our broad-bandwidth data, and cleaned down to $0.01$ mJy. A list of parameters used, and their values, are given in Table \ref{table:clean_params}. For 3C444, the imaging parameters were consistent across the L-band and C-band data, using a Briggs robust weighting parameter of 0.0 in order to keep the optimum weighting for the visibilities. For 3C320 at L-band, we did not require manual re-weighting of the individual e-MERLIN and VLA data sets relative to one another in order to view both compact and large scale structures, as would generally be required for telescopes with different sensitivities. However, since the core and hotspot emission were much more prominent in the e-MERLIN data, during the imaging we increased the weights of the longest baselines in the combined data, using a robust value of -0.8. We were consequently required to image the VLA C-band data with a robust value of $-2.0$ to match the 0.5 arcsec synthesized beam of the e-MERLIN data at L-band. 

\begin{table}
    \centering
    \begin{tabular}{l|l}
        Clean keyword & Value \\
        \hline
        \texttt{threshold} & $0.01$ mJy \\
        \texttt{psfmode} & `clark'\\
        \texttt{mode} & `mfs'\\
        \texttt{imagermode} & `csclean'\\
        \texttt{ftmachine} & `mosaic'\\
        \texttt{multiscale} & 0,2,5,10
        $\times$ beam\\
        \texttt{robust} & -0.8 -- 0.0
    \end{tabular}
    \caption{Clean parameters used in CASA to image the 3C320 and 3C444 data at L and C-band. Where different parameter values were used, we simply state the range. `Beam' refers to the synthesized beam. Most default parameters are not listed.}
    \label{table:clean_params}
\end{table}

Self-calibration was performed for the individual array data for both 3C320 and 3C444 before combination. Phase gain solutions were derived with 30 second solution intervals assuming a reference antenna near the geometric center of the array configuration. Only one iteration of amplitude self-calibration was performed after phase self-calibration, with further iterations degrading the image quality.

\subsubsection{X-ray data}
Two separate \textit{Chandra} observations of 60ks and 50ks for 3C320 were made, and reprocessed using the Chandra Interactive Analysis of Observations \citep[CIAO;][]{ciao} software package from the level 1 events files with \textsc{ciao} 4.7 and \textsc{caldb} 4.6.7. The \textit{chandra\_repro} pipeline was subsequently used to reprocess the data to produce new level 2 events files using standard CIAO analysis methods. For analysis, the merged events files were initially filtered to allow counts between 0.5-7 keV. The observations were then combined before spectroscopic analysis. The \textit{Chandra} observations for 3C444 presented by \cite{cros11}, which we use in Section \ref{sect:shock_measurements}, were processed in a similar way with \textsc{ciao} 4.6.1 and \textsc{caldb} 4.6.4.

The \textit{XMM-Newton} events files for 3C444 were reprocessed with the latest calibration data using \textit{XMM-Newton} \textsc{sas} v11.0.0. The pn camera data were filtered to include only single and double events (PATTERN  4), and data from the MOS cameras were filtered using the standard pattern mask (PATTERN 12). The data sets were also filtered to remove bad pixels, bad columns, etc. We checked each events file for flares using the light curves at higher energy levels than those emitted by the sources and used good-time-interval (GTI) filtering to select data where the light curve was within 20\% of the quiescent level. We then used \texttt{evigweight} to correct the events files for vignetting. The particle background in the \textit{XMM-Newton} sources was removed using the method described by \cite{cros08}. This uses closed filter files that were processed, filtered, and weighted in the same manner as the source data sets. The closed filter data were rotated to match the source observations, and scaling factors were calculated by comparing the count rates of the pn and MOS cameras. The closed filter data were then scaled by these factors before carrying out background subtraction when generating profiles and spectra.

We generated images for the \textit{XMM-Newton} sources using the method described by \cite{cros08}. An image was extracted for each of the three EPIC cameras using \texttt{evselect}. The MOS images were then scaled to make their sensitivity equivalent to the pn camera image so that there would be no chip-gap artifacts when the three images were combined. We generated exposure maps for each camera using \texttt{eexpmap}, which were used to correct for the chip gaps, but not for vignetting as this leads to incorrect scaling of the particle background that dominates at large radii. The resulting image is therefore not vignetting corrected; it is purely pictorial and not used in any subsequent analysis.

The \textit{Chandra} and \textit{XMM-Newton} images of the large scale environments surrounding 3C230 and 3C444, respectively, are shown in Figure \ref{figure:clusters}. These figures have been produced using the \textit{Chandra} Imaging and Plotting System (\textsc{chips}).

\section{Analysis}
\label{sect:analysis}

\begin{figure}
    \centering
    \subfloat{\includegraphics[scale=0.4, trim={24cm 5cm 24cm 3cm},clip]{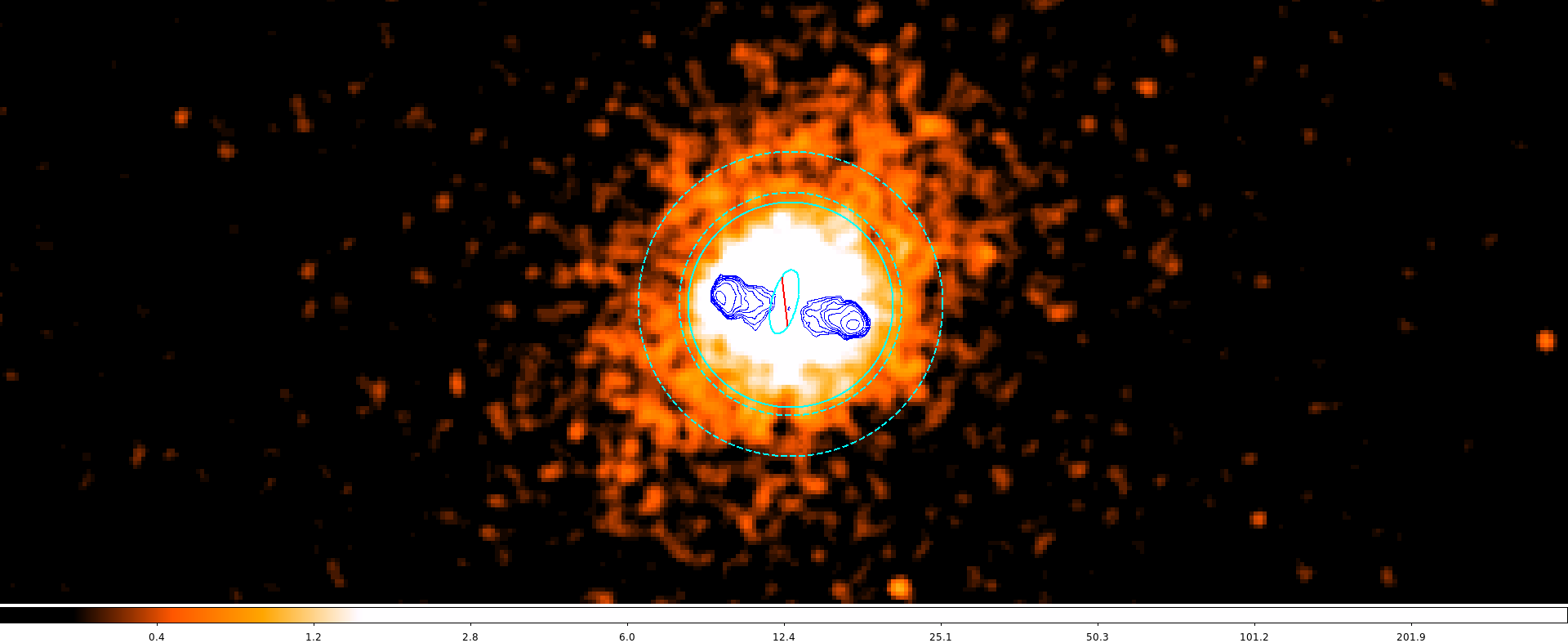}}\\
    \subfloat{\includegraphics[scale=0.32, trim={21cm 2cm 22cm 0cm},clip]{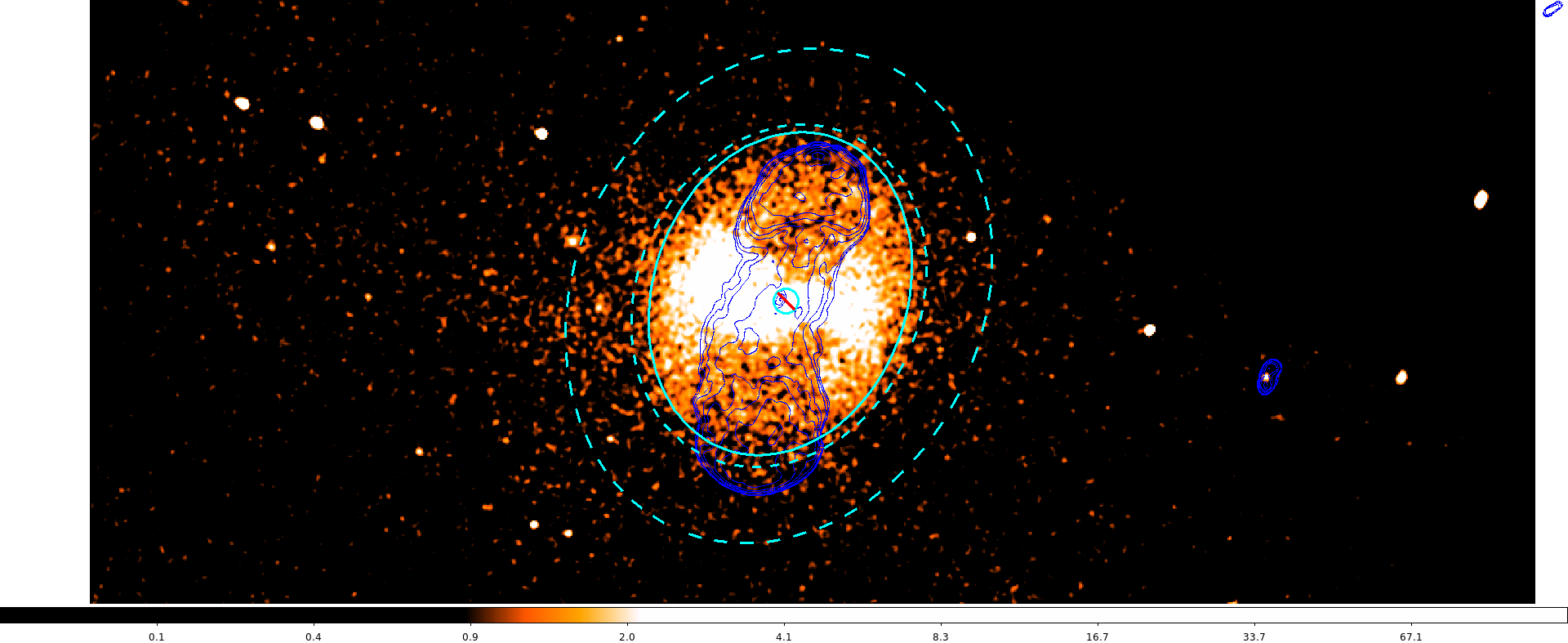}}
    \caption{\textit{Chandra} X-ray maps for 3C320 (top) and 3C444 (bottom), with radio source contours in blue from our VLA 1.4 GHz radio observations overlaid, showing regions (cyan) used to measure the ICM shock. Solid cyan lines represent the region used to extract spectra for the shock, and dashed cyan annuli represent the region used for background subtraction. The central ellipse and circle with a diagonal red line show the central AGN-related components masked out in our analysis for 3C320 and 3C444 respectively.}
    \label{fig:shock_measurements}
\end{figure}
\begin{figure}
    \centering
    \subfloat{\includegraphics[scale=0.46, trim={1.1cm 1cm 2.5cm 12cm},clip]{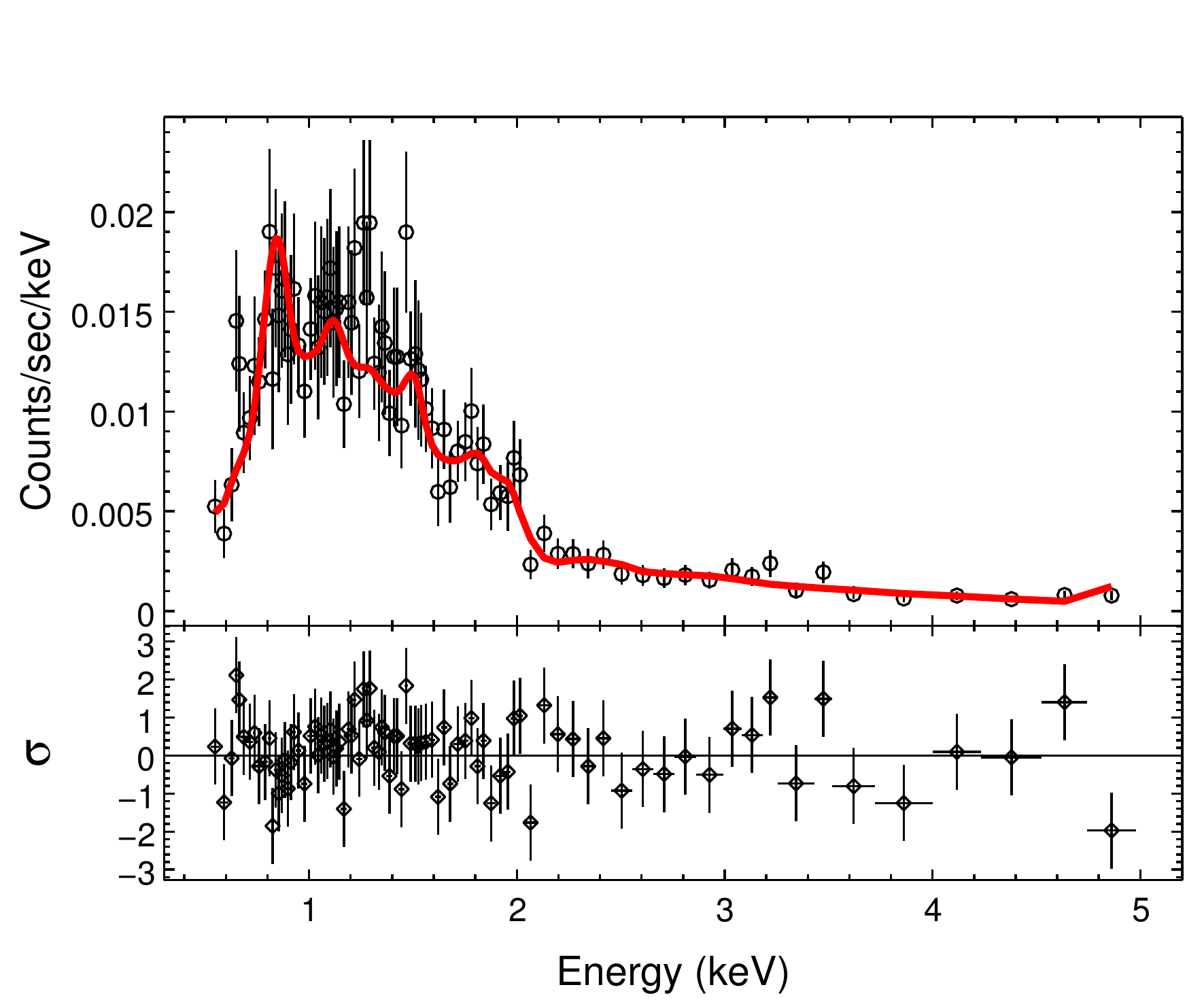}}\\
    \subfloat{\includegraphics[scale=0.46, trim={1.1cm 1cm 2.5cm 12cm},clip]{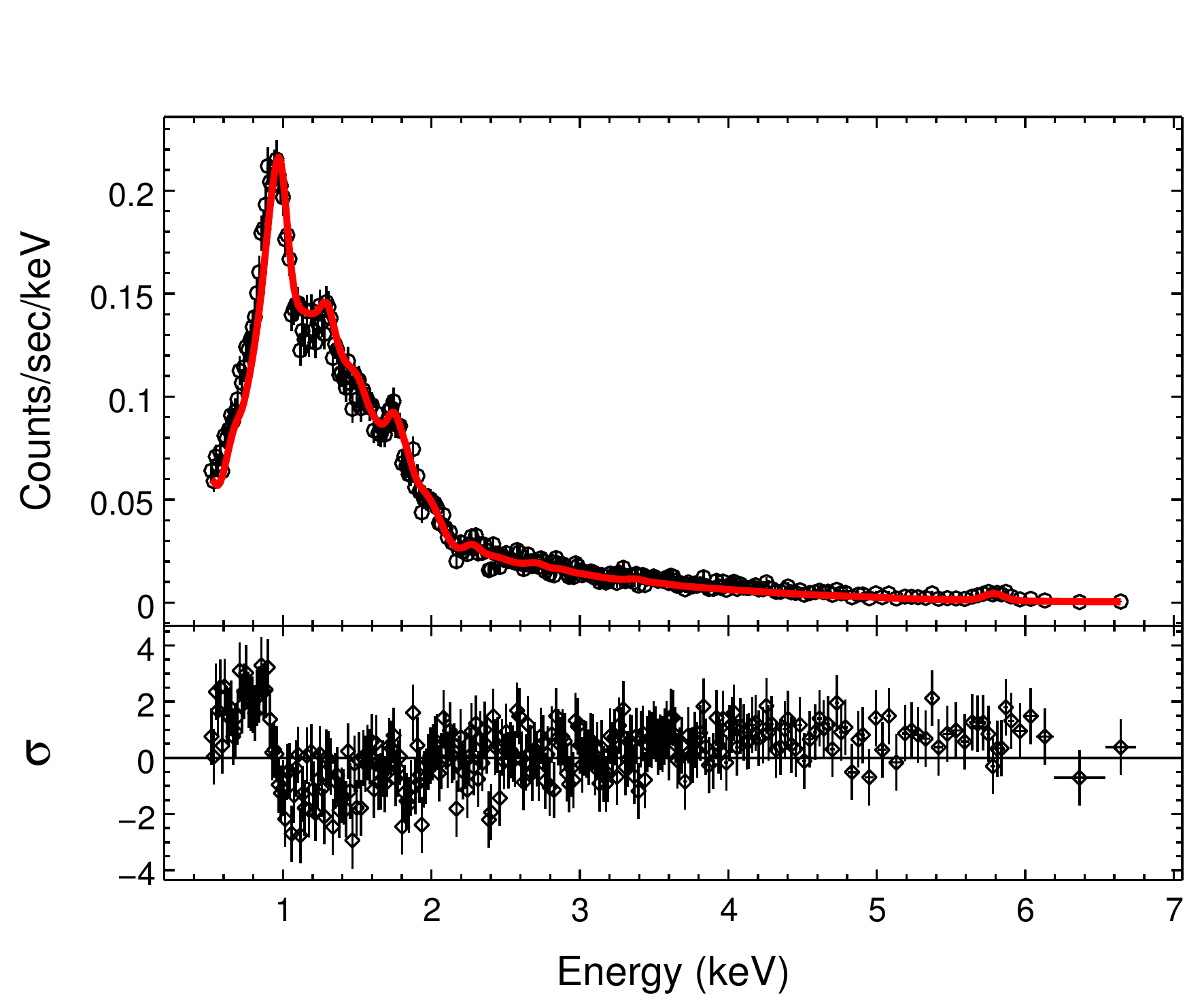}}
    \caption{\textit{Chandra} X-ray spectra for the background-subtracted shock region for 3C320 (0.5-5 keV;top) and 3C444 (0.5-7 keV;bottom) as defined in Figure \ref{fig:shock_measurements}. The red solid line indicates the best fit absorbed thermal APEC model to the data. The lower panel in each figure indicates the reduced $\chi^2$ statistic at each data point. }
    \label{fig:shock_fits}
\end{figure}
In this section we present the methods used to determine the spectral
and dynamical ages of 3C320 and 3C444. We first describe our methods
of obtaining constraints on physical parameters that are required for
the spectral ageing and dynamical age calculations; these are the ICM
shock properties, the immediate and larger-scale environments and the
lobe magnetic field strength of the sources. We then proceed to describe our methods to obtain spectral ages, followed by modelling the dynamics of both sources to obtain their dynamical ages.
\subsection{Source physical properties}
\subsubsection{Shock measurements} \label{sect:shock_measurements}
To determine the physical properties of the shocked X-ray emitting ICM
surrounding 3C320 and 3C444, we extracted the spectrum from the
calibrated \textit{Chandra} data (we used the archival
\textit{Chandra} observations analysed by \citealt{cros11} rather than
the \textit{XMM-Newton} observations for 3C444 as presented in Figure
\ref{figure:clusters}, to better constrain the shock with higher
angular resolution). Count rate files were extracted from the merged
events files on two regions of interest; the centrally shocked medium,
defined as the central region bounded by a visually clear surface
brightness jump, and a background region defined as an annulus
directly around the shock (see Figure \ref{fig:shock_measurements}).
We used the {\sc ciao} software package to obtain spectra for these regions
and {\sc Sherpa} for spectral fitting.

Bad data were removed using the \texttt{ignore\_bad()} command, removing bins based on bad data flags. The data were then filtered in the energy range 0.5-5.0 keV and 0.5-7.0 keV for 3C320 and 3C444 respectively -- allowing a larger energy range for the 3C320 observations resulted in poor fits when analysing the individual lobe regions as described in Section \ref{sect:lobebfields}, and hence we keep the energy range analysed for the shock and lobe regions consistent. Background emission was subtracted based on the count rates in the immediate vicinity of the shock (dashed lines in Figure \ref{fig:shock_measurements}).

For both regions, a thermal \textit{APEC} model was fitted to the spectra, describing thermal emission from collisionally-ionized diffuse gas, and such a spectrum is expected from the X-ray emitting gas of the intra-cluster medium. A photo-electric absorption model was also added as a multiplicative component to the \textit{APEC} model to account for the absorption of X-ray photons by foreground atomic matter such as hydrogen atoms in cold gas. An added non-thermal power-law emission model was also tested for the shock region, as would be expected due to non-thermal processes such as inverse-Compton scattering from the radio source, but resulted in worse fits and also a non-detection of non-thermal emission, and hence we considered only the absorbed thermal model to describe the shock. The Galactic absorbing column density was fixed at $1.64\times 10^{20}$ atoms cm$^{-2}$ and $2.51\times 10^{20}$ atoms cm$^{-2}$  (based on previous observations) as well as the redshift at 0.342 and 0.153 and the elemental abundance at 0.3$\times$solar, for 3C320 and 3C444, respectively. The only free parameters for the thermal model were the temperature and normalisation. The best fit values for the physical parameters of the fitted models for the shocked medium are tabulated in Table \ref{table:xray_shock_fits}. The background-subtracted spectra for the shock regions and their best fit thermal models (solid red line) are shown in Figure \ref{fig:shock_fits}.
\begin{table*}
\centering
 \caption{Shock physical properties for 3C320 and 3C444 based on our \textit{Chandra} observations. All uncertainties are based on $3\sigma$ measurement errors, with the exception of the volume uncertainties for 3C444, calculated using geometrical arguments. Note that we do not have volume uncertainties due to projection effects for 3C320 due to its circular shock region. See Section \ref{sect:shock_measurements} for more details.}  
  \begin{tabular}{lccccc}
  \hline 
  Region & $\chi^2_{\text{reduced}}$ & $kT$ (keV) & Norm ($\times 10^{-4}$) & $n_e$ ($\times 10^2$(cm$^{-3}$)) & Volume $\times 10^{70}$cm$^{3}$\\
  \hline
  3C320 & $0.841$ & $3.488^{+0.700}_{-0.542}$ & $2.138^{+0.176}_{-0.164}$ & $1.640\pm 0.065$ & $2.033$ \\[2pt] 
  3C444 & $0.960$ & $2.845^{+0.150}_{-0.140}$ & $15.534^{+0.278}_{-0.278}$ & $0.680^{+0.537}_{-0.154}$ & $18.831^{+7.800}_{-0.000}$ \\[2pt]
  \hline
 \end{tabular}
\label{table:xray_shock_fits}
\end{table*}
The electron density $n_e$ is directly calculated using the fitted normalisation to the thermal APEC model;
\begin{ceqn}
\begin{align}
\centering
\text{norm}= \frac{10^{-14}}{4\pi\left(D_A\left(1+z\right)\right)^2}\int n_en_HdV
\label{equation:norm}
\end{align}
\end{ceqn}
where $D_A$ is the angular diameter distance, $z$ is the redshift, $n_H$ is the number density of hydrogen atoms. We compute the integral in Equation \ref{equation:norm} analytically since we have determined volume-averaged parameters and we can make the assumption that $n_e$/$n_H\approx1.2$. Hence, we only need to compute $\int dV=V$, for the shocked regions shown in Figure \ref{fig:shock_measurements}. 

Region volumes were initially determined using projected angular sizes
-- the projected shocked region for 3C320 was fitted as a circle,
leading to a calculation of the volume of a sphere with the radius
defined as shown in Figure \ref{fig:shock_measurements}.
For the more complex 3C444, the projected emission is
  in the form of an ellipsoid, and hence the true physical volume may
  be different from the projected volume depending on the orientation angle to the line of sight. We assume the physical depth to be equal to the projected minor axis (i.e. a prolate spheroid), as typically modelled for the shock in dynamical models \citep[e.g.][]{hard18}. To determine the geometrical uncertainties due to orientation, we determine plausible upper and lower limits of the major axis. The projected major axis can never be larger than the physical major axis, and hence the lower limit volume of the shocked shell is obtained when we use the projected major axis. The shocked shell may then be any orientation angle to the line of sight, but since we observe double radio lobes which imply that the source is at a large angle to the line of sight, we place a lower limit on the orientation angle of 45$^{\circ}$ to the line of sight. The upper limit volume uses the upper limit major axis, which is the projected length divided by cos($45^{\circ}$). For both sources, the projected lobe volumes were subtracted from the shock volume (orientation effects of the lobe sizes on the lobe-subtracted shock volume are considered in Section \ref{sect:modelling}), as we only consider emission from the shocked ambient medium. We also subtracted the volume of a `cold bar' of emission seen over the core of 3C320, and the central AGN point source of 3C444 (masked out regions with a red diagonal line in Figure \ref{fig:shock_measurements}). Electron densities, shock volumes and fitted model temperatures for 3C320 and 3C444 are given in Table \ref{table:xray_shock_fits}.
\subsubsection{Central environment}
\label{sect:centralenv}
Here we describe the extraction of the environmental properties of 3C320 and 3C444 from our \textit{Chandra} and \textit{XMM-Newton} X-ray observations of the ICM, respectively. In particular, for the environment input parameters of the analytic model which will later be used to determine dynamical ages (Section \ref{sect:modelling}), we require the ICM thermodynamic properties at the time the radio source switched on. Assuming the hot gas from the ICM can be well described by an isothermal beta model \citep{beta}, where
\begin{ceqn}
\begin{equation}
p = p_0\left[1+\left(\frac{r}{r_c}\right)^2\right]^{-\frac{3\beta}{2}},
\end{equation}
\end{ceqn}
for the analytic model we require the constant environmental
temperature ($kT$), central pressure ($p_0$), core radius ($r_c$) of
the ICM and the beta index ($\beta$), at $t=0$. Directly fitting
emissivity models to the integrated instantaneous X-ray emission
(Figure \ref{figure:clusters}) would not necessarily translate to the
true thermodynamic state of the ICM at the time the radio source
switched on: shock heating by the radio source occurs throughout the
source lifetime, as expected since this is suggested to be the main
mode of feedback for cluster-centre radio galaxies, essentially
removing information on the past history of the central parts of the
ICM. We therefore fitted an isothermal beta model only to the outer
un-shocked regions of the ICM, and extrapolated the best fit model
down to the central region, which we assume will give estimates of the
central density and pressure of the ICM in its undisturbed state.

\begin{figure}
    \centering
    \subfloat{\includegraphics[scale=0.40, trim={3.4cm 0 0 0},clip]{{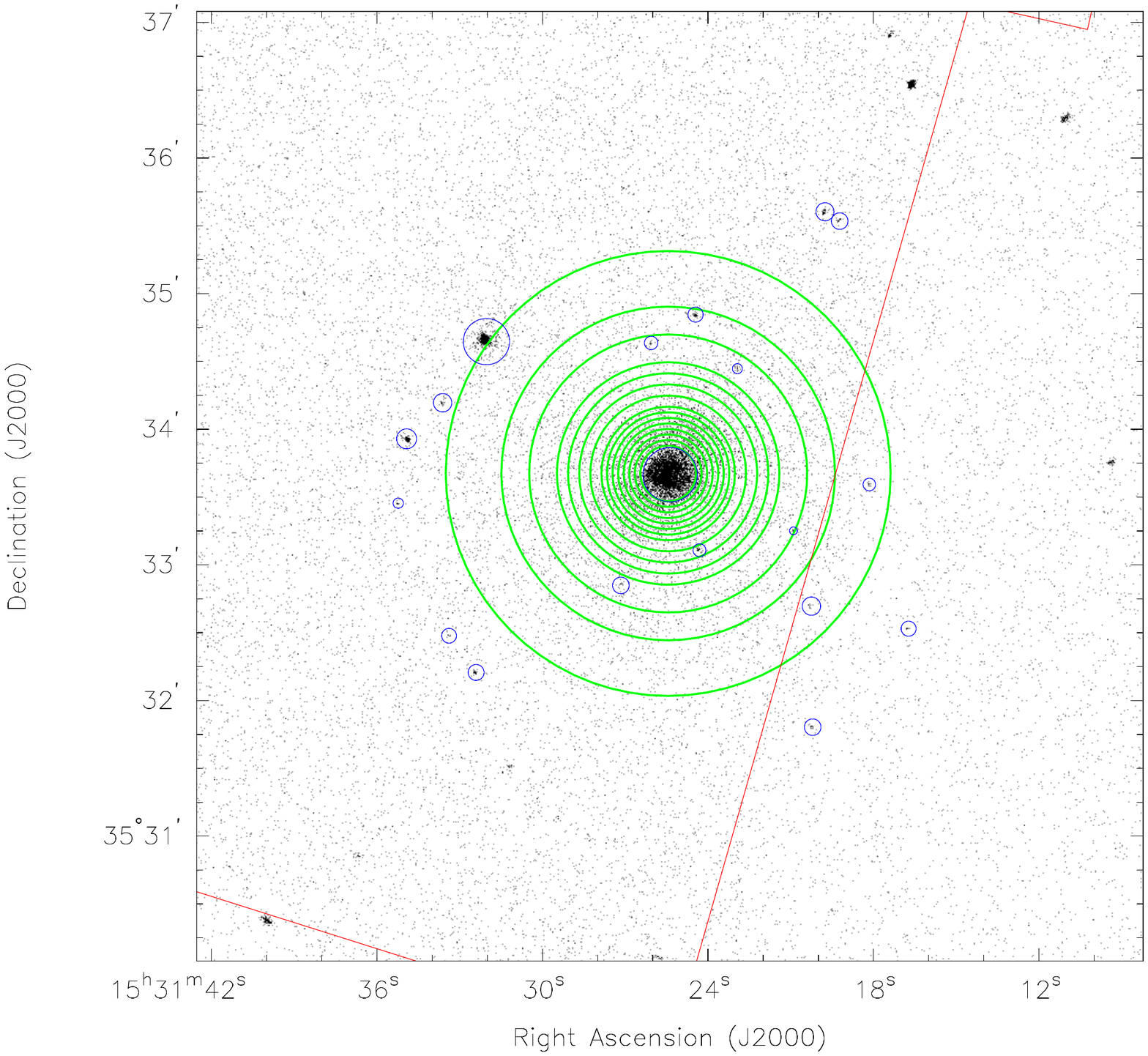}}}\\
    \subfloat{\includegraphics[scale=0.35, trim={1.8cm 0 0 0},clip]{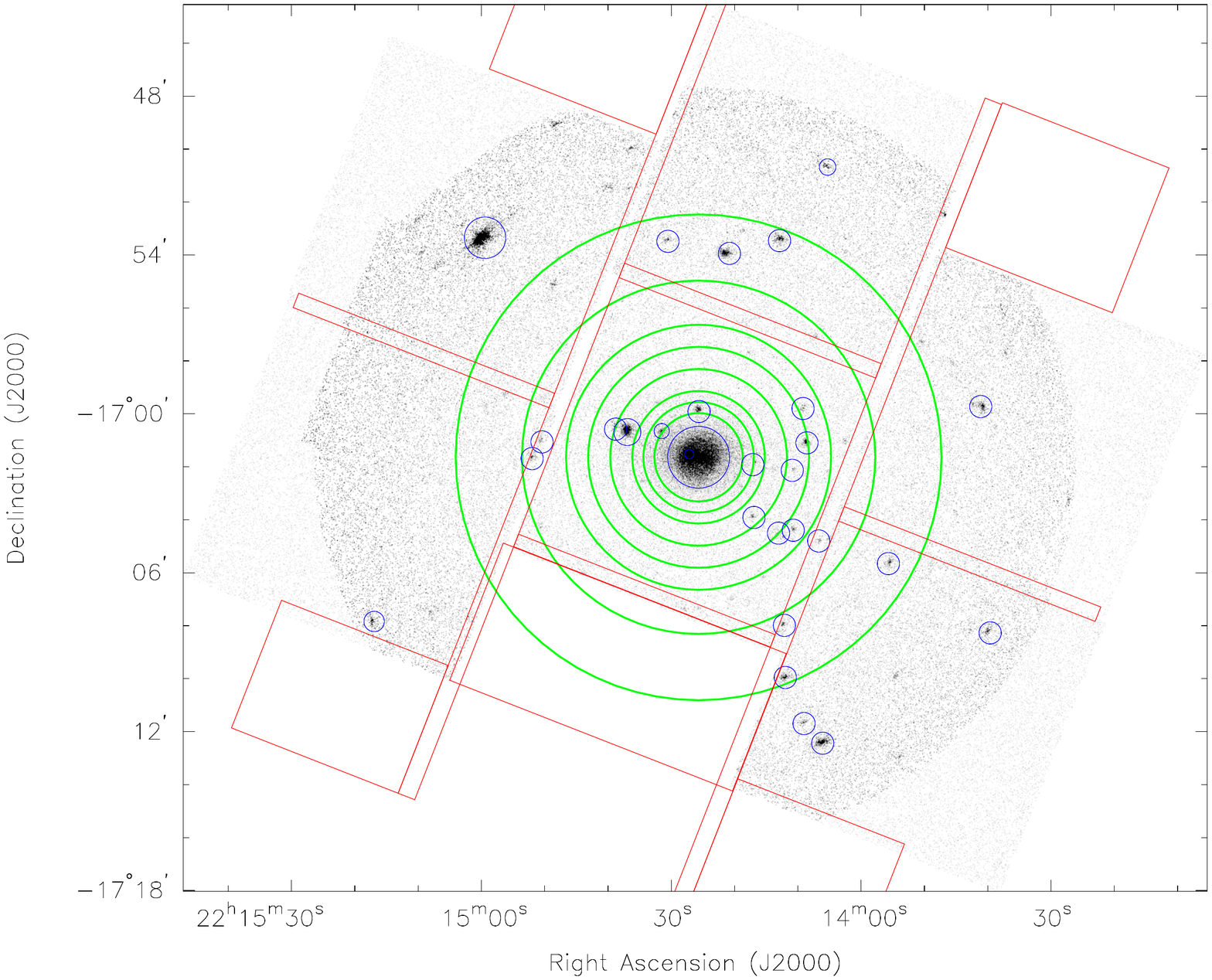}}
    \caption{Concentric annuli (green) used to extract radial profiles from the X-ray observations of the ICM surrounding 3C320 (top) and 3C444 (bottom). Masked regions for point sources and the central shock are shown in blue. Regions omitted due to chip gaps are shown as red polygons. For clarity, the region to the right of the vertical red line for 3C320 (top) marks a chip gap.}
    \label{fig:annuli}
\end{figure}

As X-ray surface brightness scales with the square of the particle density, we directly obtained surface brightness profiles to the X-ray emission of the un-shocked media of 3C320 and 3C444 using tools from the \textsc{funtools} library \citep{funtools}, defining concentric annuli outside the central shock. The first annulus was chosen to be directly outside the clear central shock region, with the annuli increasing in width out to the extent of the ICM, shown in Figure \ref{fig:annuli}. We masked chip gaps in the data (shown as red polygons), more so for the 3C444 \textit{XMM-Newton} data due to its larger field of view. 
\begin{figure}
    \centering
    \subfloat{\includegraphics[scale=0.45, trim={1cm 0cm 0 9cm},clip]{{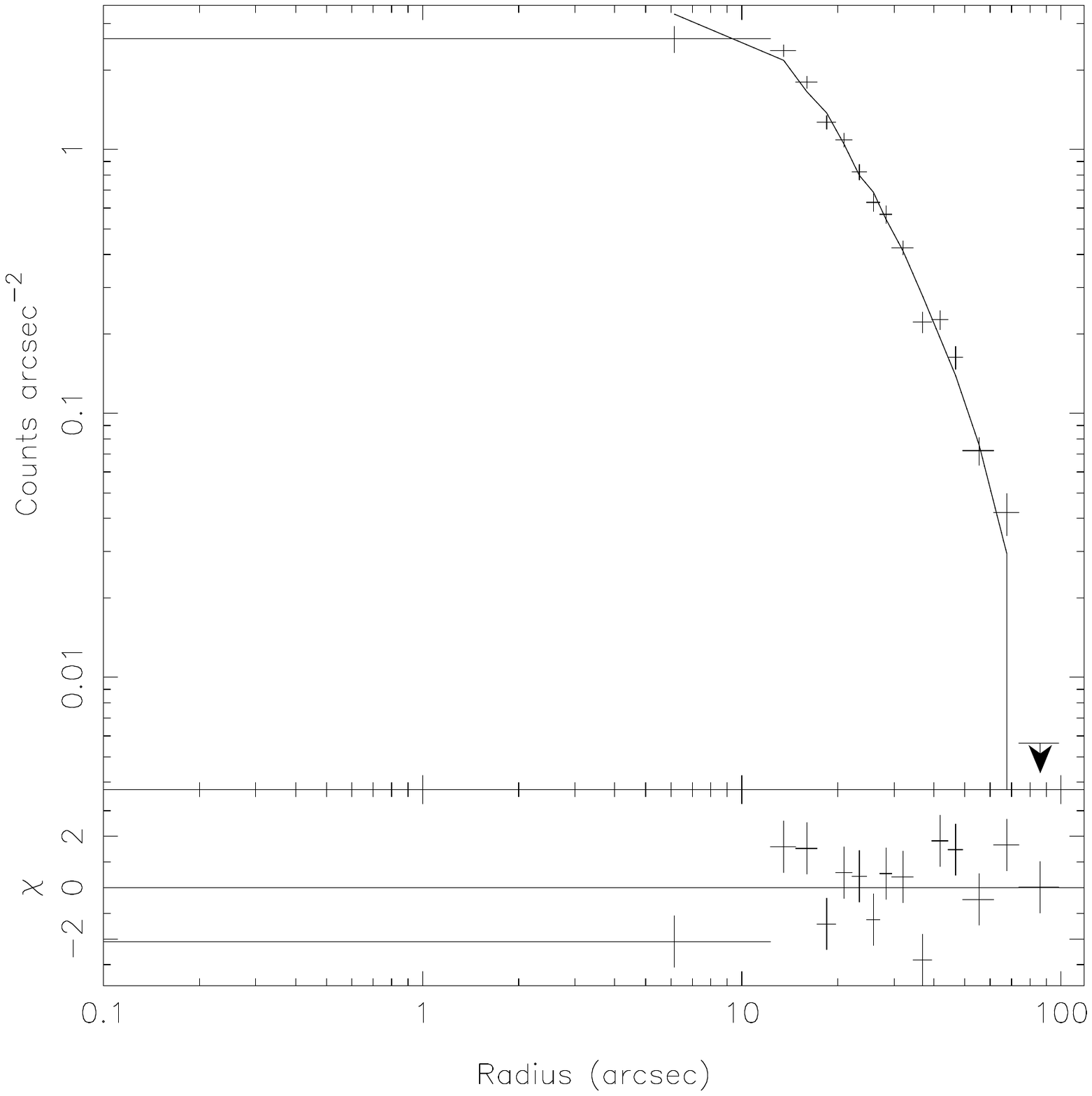}}}\\
    \subfloat{\includegraphics[scale=0.45, trim={1cm 0cm 0 9cm},clip]{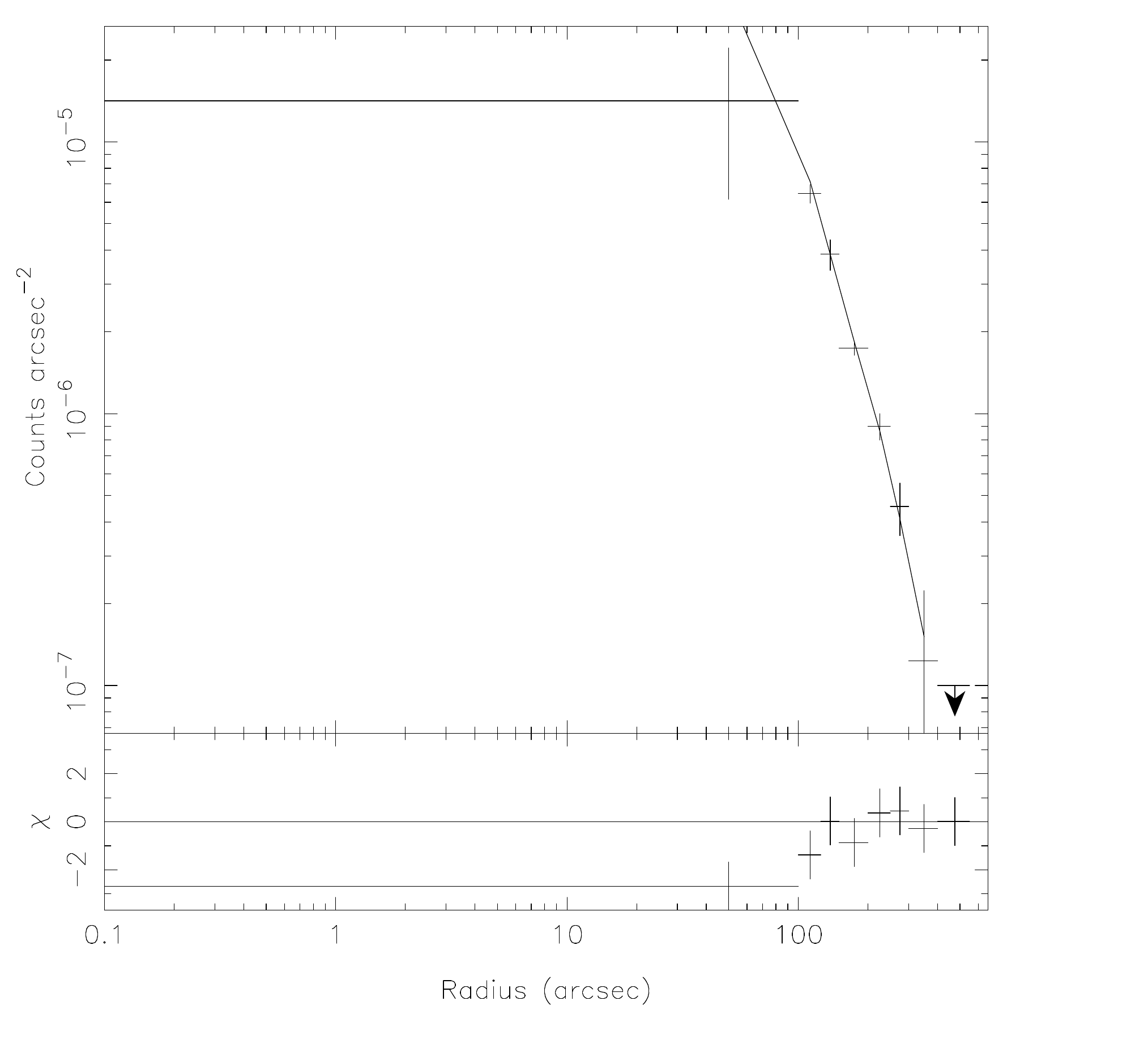}}
    \caption{Radial surface brightness profiles for the outer (un-shocked) ICM for 3C320 (top) and 3C444 (bottom). The solid lines indicate the best fit beta model, with the residual $\chi^2$ values in the lower panel of each figure.}
    \label{fig:radial_profiles}
\end{figure}
After obtaining the radial surface brightness profiles to the un-shocked environment, we used the Markov Chain Monte Carlo (MCMC) algorithm (as described by \citealt{cros08} and \citealt{ines13}) to determine the best fit beta model to the profiles  with four unknown parameters; the psf normalization, surface brightness normalization, $\beta$ and $r_c$, restricting the range of the parameter spaces to be within physically plausible values. An isothermal beta model was well fitted to the data points for the unshocked medium for both 3C320 and 3C444, with $\chi^2_{\text{red}}\sim 1$ for the best-fitting models, shown in Figure \ref{fig:radial_profiles}. It can be seen by the residuals (lower panel of the profiles) that the models are not well fitted to the first data point of the profile, i.e the first annulus. This is due to the fact that most of the area of the first annulus is masked by the central shock, and hence any measurement should not be used as a `true' measurement -- we arbitrarily increased the count rate error for the first data point of the 3C444 \textit{XMM-Newton} data to reflect this, essentially removing it as a measurement from the fit.

To obtain a value for the isothermal cluster temperature, we fitted a thermal APEC model to the \textit{Chandra} and \textit{XMM-Newton} X-ray emission spectra (between 0.5-5 keV and 0.5-7 keV for 3C320 and 3C444, respectively), masking out only the central shock, using the CIAO software package, giving best-fit temperatures of $2.77^{+2.26}_{-0.85}$ keV and $2.33^{+0.10}_{-0.10}$ keV for 3C320 and 3C444, respectively. Given the central density output by the beta model and the fitted ICM temperature, we then use the ideal gas equation to determine the central pressure $p_0$. The physical environmental properties we derive are then used as input to the analytic model, used to model the evolution of 3C320 and 3C444, described in Section \ref{sect:modelling}. Though we have outlined some of the assumptions used in this method in predicting the past state of the ICM, which may deviate from the true ICM properties in reality, our methods are physically motivated and represent the best possible approach with the available data.
\subsubsection{Lobe magnetic field strength}\label{sect:lobebfields}
As stated in Equation \ref{equation:specage}, the calculation of a
spectral age requires a magnetic field strength, as one of the
important parameters that govern the strength of radiative losses for
a given electron energy, and the magnetic field strength is therefore
an input parameter for the spectral age model fitting. As discussed in
Section \ref{sec:equipartition}, equipartition estimates are likely to
overestimate the true magnetic field, and inverse-Compton constraints
are desirable. We used our \textit{Chandra} X-ray observations of the ICM surrounding 3C320 to search for inverse-Compton emission from the lobes, and a similar method was used to determine the field strength for 3C444 by \cite{cros11} -- we simply state their results here. We expect the dominant source of emission in our observations at the location of the radio source to be thermal bremsstrahlung rather than non-thermal inverse-Compton emission directly from the lobes -- the shock region is physically larger than the lobes and includes the integrated line of sight emission through the cluster, consisting of thermally emitting material. Moreover, it is difficult to entirely remove thermal foreground cluster emission in front of the lobes using only background subtraction, and hence any spectral-based fitting of non-thermal models on large regions is expected to be contaminated with thermal emission \citep[e.g.][]{hard10}. We used the \textit{Sherpa} application on the \textsc{ciao} software to fit thermal and power-law models to elliptical regions at the location of both radio lobes in the \textit{Chandra} observations. As the power-law model contains additional free parameters (the photon power law index $\Gamma$ and the corresponding normalisation), we fixed the lobe temperature to that of the fitted temperature of the shocked medium found in Section \ref{sect:shock_measurements}, reducing the number of free parameters while keeping our analysis consistent. Initial fits resulted in $\Gamma$ being relatively unconstrained by large errors. Subsequent fitting with a fixed power-law index resulted in consistent values for the power-law and thermal normalisations and the fitted $\chi^2_{\text{reduced}}$ for $1.5\leqslant\Gamma\leqslant2.0$. We therefore fixed the power-law index at $\Gamma=1.75$ to improve the fitting (consistent with typically measured spectral slopes of radio lobes of $\alpha\sim 0.7$, as $\Gamma=\alpha+1$). The results for the emission model fits for both lobe regions in the shock are given in Table \ref{table:lobe_pl_fits}.

\begin{table}
\centering
 
  \begin{tabular}{lcccc}
  \hline
  Lobe & $\chi^2_{\text{red}}$ & PL Norm & Thermal Norm & Flux density (nJy)\\
  \hline
  East & $0.973$ & $0.000^{+0.0022681}$ & $0.420^{+0.050}_{ -0.176}$ & < 2.268 \\ 
  West & $0.653$ & $0.000771^{+0.00522}$ & $0.531^{+0.113}_{ -0.531}$ & < 6.000\\
  \hline
 \end{tabular}
 \caption{Emission model fits to the ICM, spatially coincident with the lobes of 3C320. Regions modelled with an absorbed power-law model plus thermal APEC model with fixed values of $\gamma$ = 1.75, thermal abundance at $0.3\times$ solar and the temperature at the fitted shocked temperature of 3.488 keV. $\gamma$ is defined in the sense $A(E) \sim E^{-\gamma}$. Note that the power-law normalisation was set such that it returns the 1 keV flux density in units of $\mu$Jy. Flux densities are given as 3$\sigma$ upper limits.}  
\label{table:lobe_pl_fits}
\end{table}
As detailed in Table \ref{table:lobe_pl_fits}, we were not able to detect  a non-thermal power-law component in the lobe regions of the cluster emission. A 3$\sigma$ upper limit flux density of inverse-Compton emission was obtained, giving flux densities of 2.268 nJy and 6 nJy for the eastern and western lobes, respectively. These are still within the range of detected inverse-Compton fluxes obtained for the lobes of FR-II galaxies \citep[][]{erlu06,good08}. 

We used the \textsc{synch} code \citep{hard98} to determine the equipartition lobe magnetic field based on the observed radio flux densities measured with our VLA observations, giving a field strength of $B_{eq} \approx 4.4$ nT for both lobes of 3C320 (assuming $\alpha_{inj}=0.60$ for the low-frequency radio data as the expected value between the theoretical and steeper values from recent studies -- the calculated field strength is not largely dependent on this value). Since the code also predicts the inverse-Compton spectral energy distribution based on a fixed magnetic field, we fixed the magnetic field strength value and ran the fitting process for a range of field strengths lower than the equipartition value until the measured upper limit flux density at 1 keV was consistent with the model-predicted inverse-Compton spectral energy distribution at the same energy. The magnetic field giving this consistency was then treated as a lower-limit to the lobe magnetic field based on inverse-Compton constraints. Thus, we obtained both a lower and upper limit (from equipartition) lobe magnetic field strength for 3C320, as derived similarly for 3C444 \citep{cros11}. The derived field stength limits are given in Table \ref{table:lobe_bfield_fits}. Taking approximately averaged values, we take the field strength for both lobes to be within the range $0.50 - 4.44$ nT. For 3C444, \cite{cros11} obtain a magnetic field strength range of $0.5-1.0$ nT. We use these constraints to the lobe magnetic field to calibrate evolutionary models for 3C320 and 3C444 in Section \ref{sect:modelling}.
\begin{table}
\centering
 
  \begin{tabular}{lcccc}
  \hline
  Source & Upper limit (nT) & Lower limit (nT)\\
  \hline
  3C320; east lobe & $4.41$ & $0.60$ \\ 
  3C320; west lobe & $4.47$ & $0.35$ \\
  3C444 & $1.05$ & $0.50$ \\
  \hline
 \end{tabular}
 \caption{Derived magnetic field strengths for the lobes of 3C320 (this work) and 3C444 \citep{cros11}. The upper limits are calculated using the equipartition assumption and the lower limits are based on the upper-limit inverse-Compton measurements.}
\label{table:lobe_bfield_fits}
\end{table}
\subsection{Spectral age fitting}
\label{sect:specages}
We fit spectral ageing models to the radio lobe spectra using the {\sc brats} software package. The best fitting ages for a given input magnetic field gives the break frequency, which we later use to re-determine spectral ages after constraining the magnetic field strengths from our limits using our analytic modelling. Essentially, for a given age and magnetic field, the model fluxes are computed numerically for each region in the radio map and quantitatively compared with the observed flux for the same regions at the frequency of the map, producing a set of $\chi^2$ values. The best-fitting ages for each of the predefined and resolved regions in the map are then output (see \citetalias{harwood13} for further details).

The currently favoured models are the Jaffe-Perola \citep{jpmodel} and Tribble \citep{tribblemodel,hard13} models of spectral ageing, and we limit our analysis to these models (see \citetalias{harwood13,harwood15} and \citealt{harw17} for results on comparisons between models). To determine the source flux density $S_{\nu,i}$ at each intermediate frequency $i$, we produced sub-band images from the broad-bandwidth observations of 3C320 and 3C444 seen in Figure \ref{figure:radio_obs}. We imaged each spectral window (our L-band data has 16 and our C-band data has 32), producing sub-band maps with 64 MHz bandwidth at L-band and 128 MHz at C-band. Note that for 3C320 at L-band, we only considered the central half of the VLA bandwidth (8 spectral windows with 64 MHz bandwidth) to match the bandwidth covered by e-MERLIN (512 MHz). A lack of sensitivity to small-scale structures drove our need to then combine spectral windows to produce only three maps at L-band for 3C320. Moreover, we only used the first 20/32 spectral windows at C-band, due to the fact that the most diffuse structures that are detected at L-band are not detected at these higher frequencies. We did not require a reduction of useable bandwidth for our 3C444 data.

We convolved each of the sub-band maps across both L and C-band with a Gaussian beam of the same shape and size, and hence, across the entire frequency range of our observations, we are limited to the resolution of the lowest frequency data. We used a circular beam of FWHM $0.8$ arcsec for 3C320 and 2 arcsec for 3C444. Since the imaging and the phase self-calibration processes described in Section \ref{sect:observations} can introduce small spatial shifts in the maps, we performed a pixel alignment procedure by fitting a Gaussian to the radio core in all of the sub-band maps using \textsc{casa}. Using the mean pixel co-ordinates of the core, we aligned all the maps where the core was offset from this reference position. These sub-band maps were directly used as input to {\sc brats}, which determined the lobe flux density from each frequency/map to produce a spectrum. 
\begin{figure}
    \centering
    \subfloat{\includegraphics[scale=0.4]{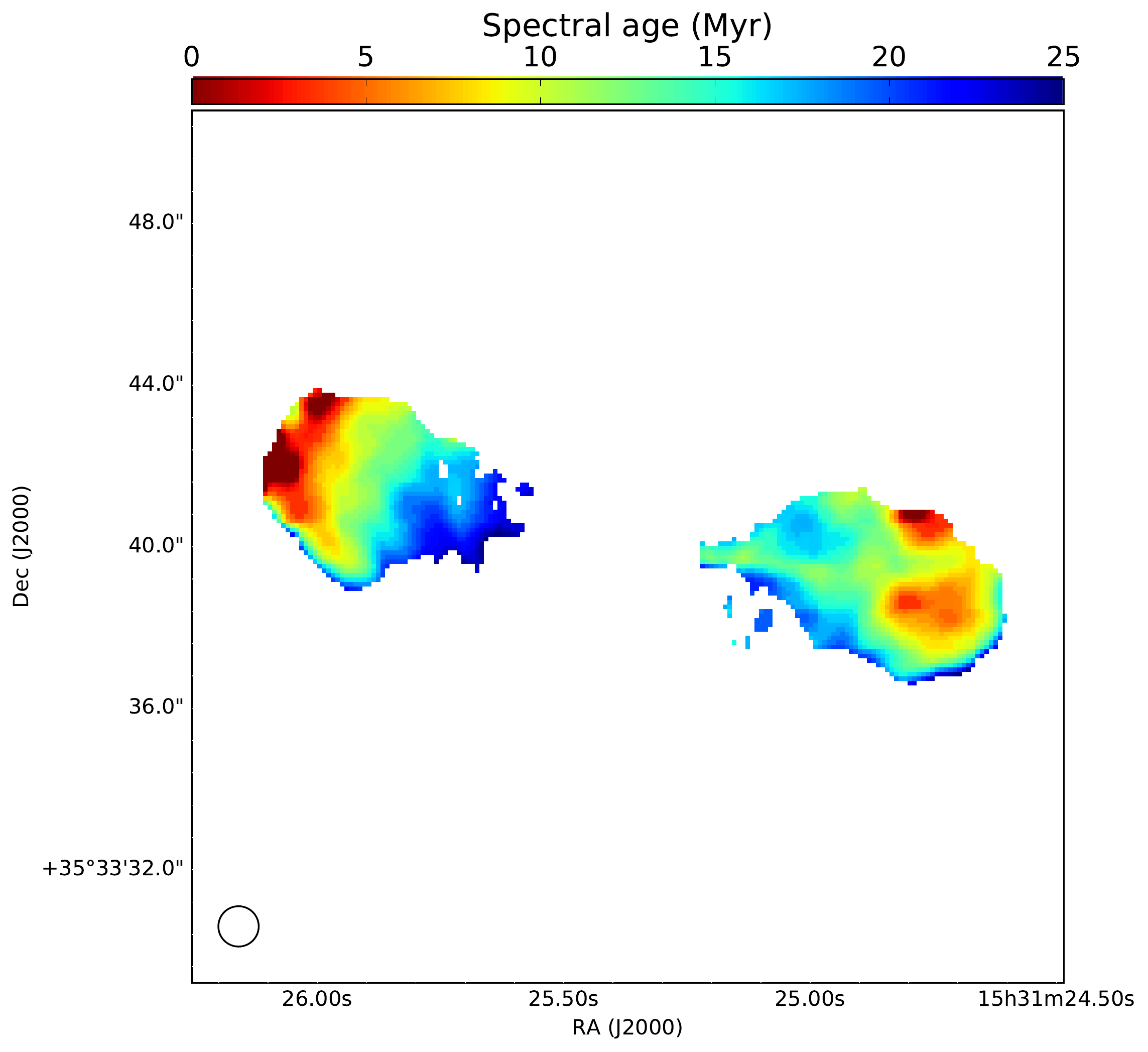}}\\
    \subfloat{\includegraphics[scale=0.4]{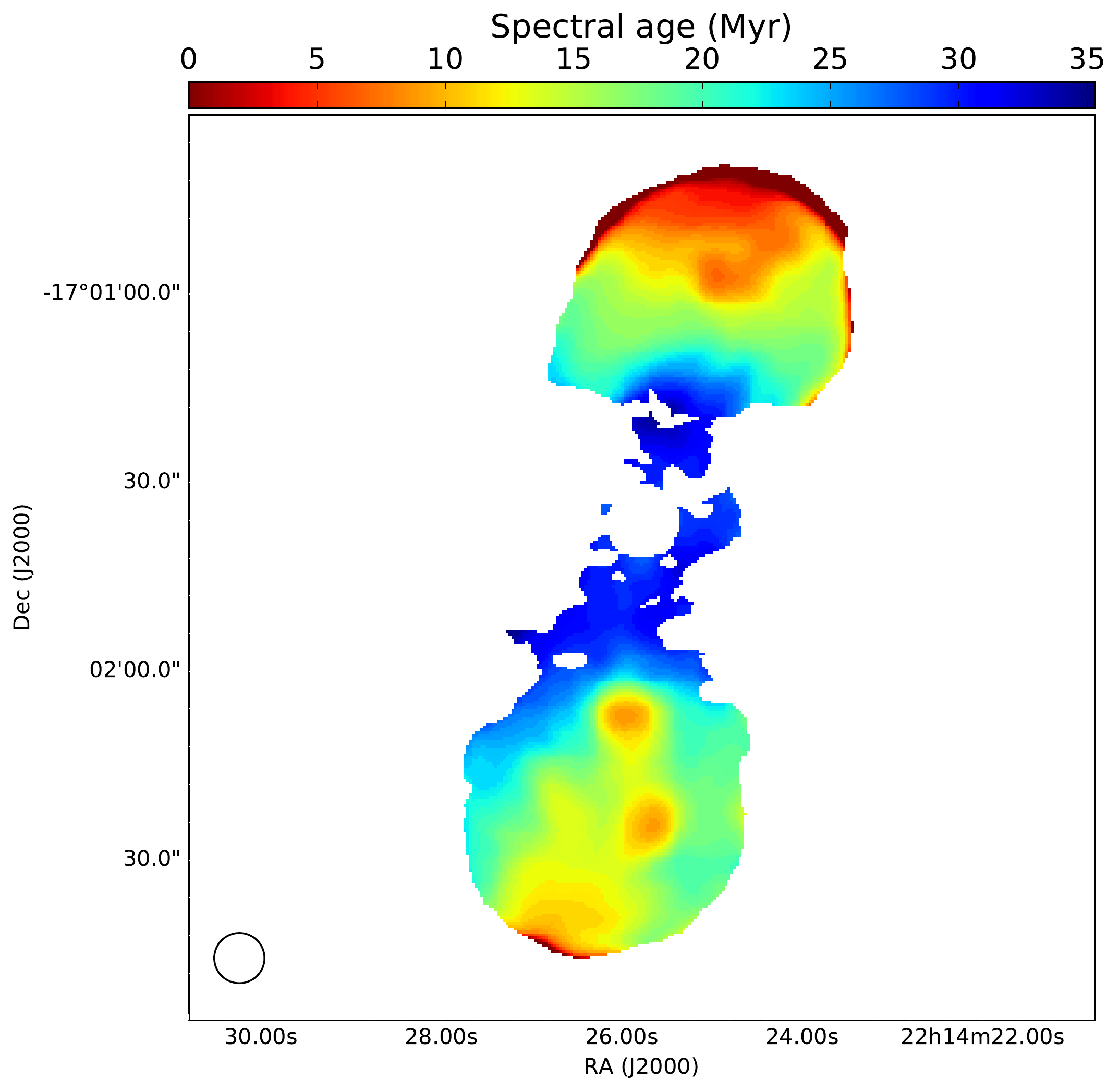}}
    \caption{Spectral ageing maps produced by {\sc brats} of 3C320 (top) and 3C444 (bottom). The colour wedge shows the ages corresponding to the colour scale. Beam sizes of 0.5 arcsec (top) and 2.0 arcsec (bottom) are shown on the lower left corner. Input injection indices are 0.6 (top) and 0.65 (bottom), while the input lobe magnetic field strengths for these maps are 1 nT (top) and 0.74 nT (bottom). The maximum spectral ages for 3C320 and 3C444 in these fits are 25.10 Myrs and 30.10 Myrs, respectively (see Table \ref{table:specages3C320}).}
    \label{fig:specagemaps}
\end{figure}
\begin{table}

\centering
 \caption{Best-fit maximum spectral ages for 3C320 and 3C444 for two ageing models. Values are given at three magnetic field strength values ranging from the lower to upper measured limits. Note that the errors quoted are only statistical errors on model fits, and are therefore represent lower limits to the true uncertainty in ages.} 
 \begin{tabular}{cccc}
  \hline 
  \multicolumn{1}{c}{\textbf{Model}} & \multicolumn{2}{c}{\hspace{1.5cm}\textbf{Spectral age (Myr)}}\\ & 0.5 nT&1.0 nT&4.4 nT \\ [5pt] \hline 
  3C320 (JP) & 35.48$^{+0.74}_{-0.82}$ & 21.98$^{+0.32}_{-0.63}$ & 3.05$^{+0.08}_{-0.07}$ \\[10pt] 
  3C320 (Tribble) & 42.90$^{+0.44}_{-1.99}$ & 25.10$^{+0.31}_{-0.23}$ & 3.78$^{+0.03}_{-0.10}$  \\[5pt]
  \hline
  & 0.5nT & 0.7nT & 1.0 nT \\[5pt] \hline 
    3C444 (JP) & $38.00^{+0.28}_{-0.20}$ & $27.48^{+0.06}_{-0.50}$ & $18.48^{+0.06}_{-0.13}$ \\[10pt] 
  3C444 (Tribble) & 42.90$^{+0.09}_{-1.03}$ & 30.10$^{+0.10}_{-0.23}$ & 20.86$^{+0.05}_{-0.09}$ \\[5pt]
  \hline
 \end{tabular}
\label{table:specages3C320}
\end{table}
The unknown physical parameters that are required by {\sc brats} for the spectral age calculation are the lobe magnetic field strength and the low frequency particle injection index. In Section \ref{sect:lobebfields} we obtained inverse-Compton and equipartition-based lower and upper limits, respectively, to the true lobe field, resulting in $0.5\text{nT}\leq B_{\text{3C320}}\leq 4.4\text{nT}$ and $0.5\text{nT}\leq B_{\text{3C444}}\leq 1.05 \text{nT}$, the latter obtained by \cite{cros11}. A factor of two difference in magnetic field will also result in a factor of two difference in the spectral age at a fixed break frequency (as $B\sim B_{CMB}$ -- see Equation \ref{equation:specage}), and hence we ran the spectral age fitting assuming three values for the magnetic field strength within the allowed limits. To determine the injection indices, we use the {\sc brats} command \texttt{findinject}, which fits for the low-frequency power law of the spectra using spectral ageing models. The injection index fitting was performed for 3C320 and 3C444 using magnetic field strengths at $4.44\times 10^{-9}$ T, $1.00\times 10^{-9}$ and $0.50\times 10^{-9}$ T, and $0.5\text{nT}, 0.74 \text{nT}$ and $1.05 \text{nT}$, respectively. The best-fit injection indices gave values around $0.6\leq q_{inj}\leq 0.65$ for any choice of magnetic field strength, for both sources, for both the JP and Tribble ageing models. The subsequently fitted spectral ages for all three magnetic fields and for all three ageing models are given in Table \ref{table:specages3C320}, and in Figure \ref{fig:specagemaps} we display spectral ageing maps for both sources using the Tribble model and the intermediate field strength from the aforementioned field strength limits. The best fit break frequencies based on the Tribble model for the oldest plasma detected are 1.7 GHz and 3.4 GHz for 3C320 and 3C444, respectively.

The process of spectral ageing can be clearly seen in these maps, with
a smooth variation from particles at the injection region with young
or zero age, to older populations towards the base of the lobes. More
than one region in each source other than in the hotspots is clearly
seen to have young populations of radiating particles, which may be
associated with jet knots created by the jet driving into locally
dense gas. It is important to note that we are not sensitive to to
faint regions at high frequencies -- in particular for 3C320, it can
be seen that the oldest regions that would exist in the vicinity of
the core (seen at L-band) are missing from the maps. We discuss the
implications of this in Section \ref{sect:discussion}.
We also note that the flux density errors on each
  sub-band map used to determine the spectrum are likely to be
  underestimated -- our data reduction and self-calibration on the
  broad-band data sets may correlate the flux density measurements
  between sub-band maps. A robust treatment of these errors would
  increase the errors on each flux density measurement by a factor of
  $\sim 2$ for the C-band points relative to the L-band points,
  increasing the leverage of the L-band data in fitting the spectra.
  While this would affect the fitted break frequency, we do not have
  accurate constraints on all possible systematic uncertainties, and
  so we use the standard VLA flux calibration errors (2 per cent at C-band) used as standard by {\sc brats}. While the oldest detected region in each source gives an indication of the age of that population of particles since injection at the hotspot, we do not have a true magnetic field estimate and hence the spectral ages listed in Table \ref{table:specages3C320} are only based on the assumed field strengths. In Section \ref{sect:modelling} we use an analytic model to determine the field strength constrained by our limits, which we then use to re-calculate the spectral ages with our fitted break frequencies.
\subsection{Analytic modelling} \label{sect:modelling}
In this section we describe the use of a analytic model to simulate the evolution of the lobes of 3C320 and 3C444, with the aim of determining more robust source dynamics than would be implied by instantaneous X-ray surface brightness. This will lead to robust source dynamical ages, for comparison with our spectral ages.

We use the analytic model of \citep[hereafter the analytic model;][]{hard18}, which models the evolution of a `shocked shell' that is driven by a radio lobe in a particular environment for a particular set of radio source properties (see \cite{hard18} for further details of model setup, input and output parameters). The modelled dynamics of the shocked shell can be used to approximate the physical properties of the expanding lobes as a function of its dynamical age. We use this model to compute the predicted evolution of shocked shells based on the observed properties of 3C320 and 3C444, and then compare their instantaneous measured radio lobe and shock properties against the model predicted properties as a function of the source age, leading to model dynamical ages constrained by our observations. Using the model lobe magnetic field at the dynamical ages and our fitted break frequencies from Section \ref{sect:specages}, we may determine robust spectral ages and test them against the modelled dynamical ages, where the dynamical evolution is governed by physically realistic environmental information based on our X-ray observations. The usefulness of this particular analytic model has already been tested by its application to a large sample of radio-loud AGN \citep{hard19}.
\begin{table}
\centering
 \caption{Table of input parameters used for the model runs. $\beta$ is the power-law index for the beta model profile fitted to the environment, $kT$ is the global cluster temperature fitted using an APEC model, $p_0$ is the central pressure of the ICM using the fitted beta model profile, $r_c$ is the core radius, $\alpha_{inj}$ is the fitted power-law index of the radio source using our spectral ageing analysis (Section \ref{sect:specages}), $z$ is the source redshift, $\zeta$ is the ratio of lobe energy densities in magnetic field and electrons, $Q_{jet}$ is the jet power and $\theta$ is the shell inclination angle.}
 \begin{tabular}{ccc}
  \hline
  Parameter & 3C320 & 3C444 \\
  \hline
  $\beta$ & $0.67^{+0.04}_{-0.04}$ & $0.72^{+0.02}_{-0.02}$ \\ [5pt]
  $kT$ (keV) & $2.77^{+2.26}_{-0.85}$ & $2.37^{+0.10}_{-0.10}$ \\ [5pt]
  $p_0$ (Pa) & $1.53^{+0.01}_{-0.13}\times 10^{-11}$ & $5.09^{+0.07}_{-0.10}\times 10^{-12}$\\ [5pt]
  $r_c$ (kpc) & $82.06^{+5.35}_{-5.94}$ & $145.03^{+8.97}_{-8.41}$\\ [5pt]
  $\alpha_{inj}$ & $0.6$ & $0.65$\\ [5pt]
  $z$ & $0.342$ & $0.153$\\ [5pt]
  $\zeta$ & $0 - 0.5$ & $0 - 0.1$\\ [5pt]
  $Q_{\text{jet}}$ (W) & $10^{37} - 10^{39}$ & $10^{37} - 10^{39}$\\ [5pt]
  $\theta$ ($^{\degree}$) & $45 - 90$ & $45 - 90$\\
  \hline
 \end{tabular}
 \label{table:beta_model_fits}
\end{table}
\subsubsection{Model setup}
In terms of the radio source properties, the model requires; the value of the low-frequency injection index ($\alpha_{inj}$), the source redshift ($z$), the ratio of energy density stored in the magnetic field to that in the radiating particles ($\zeta$) and the jet power $Q_{\text{jet}}$. We set $\alpha_{inj}$ equal to the values fitted by {\sc brats} in Section \ref{sect:specages}, which are close to theoretically predicted value based on particle acceleration in radio galaxies \citep{long73}. $\zeta$ can only be within the range $0-1$ -- although we can constrain the electron energy density using our radio observations, we do not have an exact value for the magnetic field strength. After running initial test models, the parameter space constrained by our observations resulted in models with $\zeta$ only in the range $0-0.5$ for 3C320 and $0-0.1$ for 3C444. Hence, we employ the model for a range of fifty $\zeta$ values with equal intervals in the range $0-0.5$ and $0-0.1$ for 3C320 and 3C444, respectively. Similarly for the unknown jet power, for each value of $\zeta$ modelled, we ran the model for a range of fifty jet powers in the physically plausible range of $10^{37}\leqslant Q_{\text{jet}}\leqslant10^{39}$W. We also ran each model for various source inclination angles ($\theta^{\degree}$) with respect to the observer, with twenty values from $45^{\degree}\leqslant\theta\leqslant 90^{\degree}$, with $90^{\degree}$ being a face-on radio galaxy.

In terms of the input environmental properties, the model requires
parameters based on either an isothermal beta model, or a universal
pressure profile \citep{arna10}. Throughout our modelling we use an
isothermal beta model, as has been used traditionally for many years
on observations of X-ray-bright clusters \citep[e.g.][]{mohr99}. For a
beta model atmosphere the analytic model requires: the beta index
$\beta$, central density $n_0$, isothermal temperature $kT$, central
pressure $p_0$ and the core radius $r_c$. We obtained the value of
these parameters, as described earlier in Section
\ref{sect:centralenv}, as estimates of the thermodynamical pre-shocked
state of the ICM surrounding 3C320 and 3C444. The values of the model
input parameters are given in Table \ref{table:beta_model_fits}.
50,000 models were run, for a range of $\zeta,Q_{jet},\theta$, each in
a fixed environment. The models were run for a total of 250 time steps
within the range 0.00001 -- 200 Myr for 3C320 and 3C444, with 50
logarithmic steps up to 0.1 Myr, and 200 linear steps thereafter in
order to sample well the intermediate and late-stage evolution of the source.

\begin{figure}
\centering
\subfloat{\includegraphics[scale=0.45, trim={0.45cm 0cm 1cm 1cm},clip]{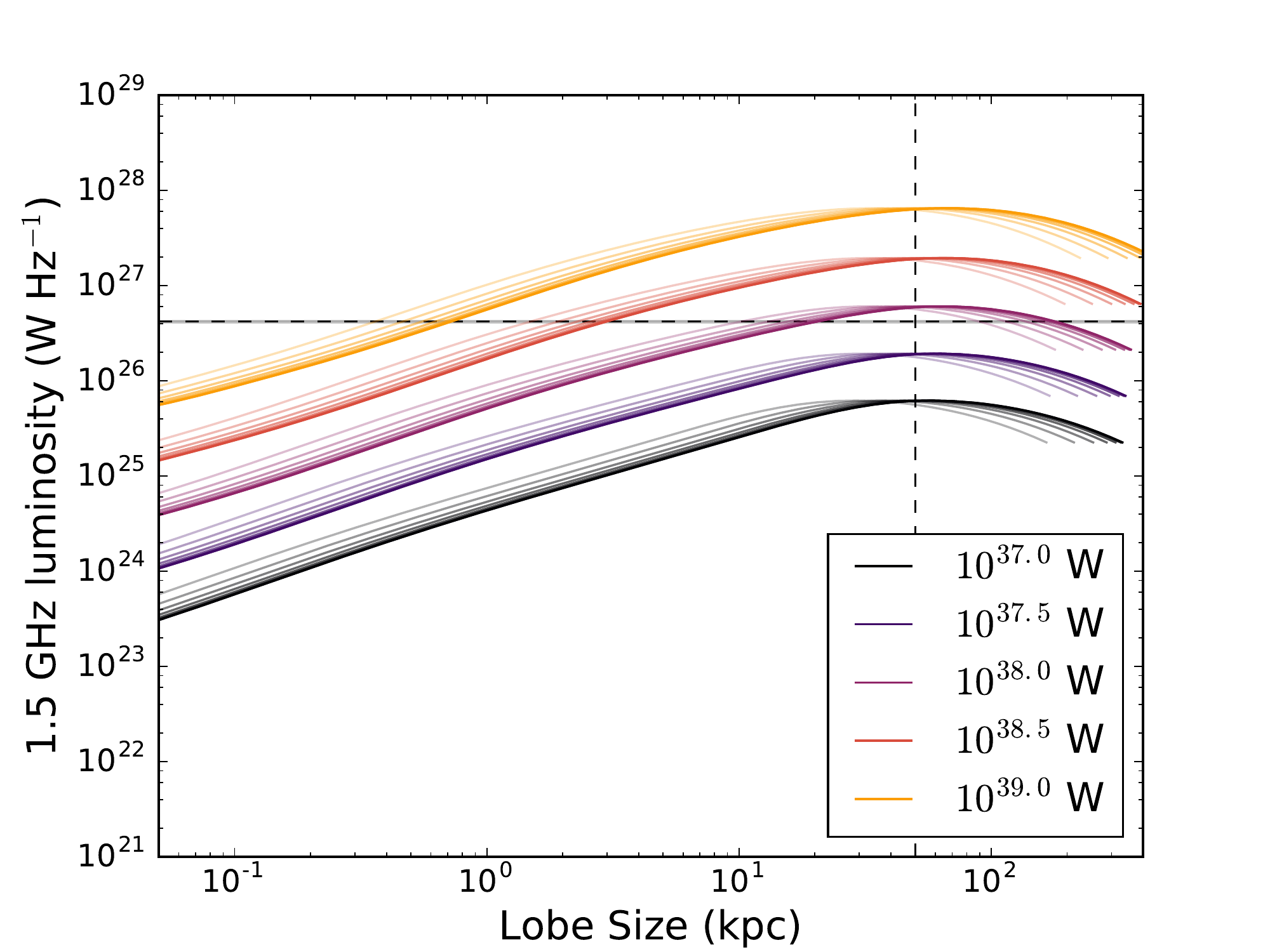}}\\
\subfloat{\includegraphics[scale=0.45, trim={0.45cm 0cm 1cm 1cm},clip]{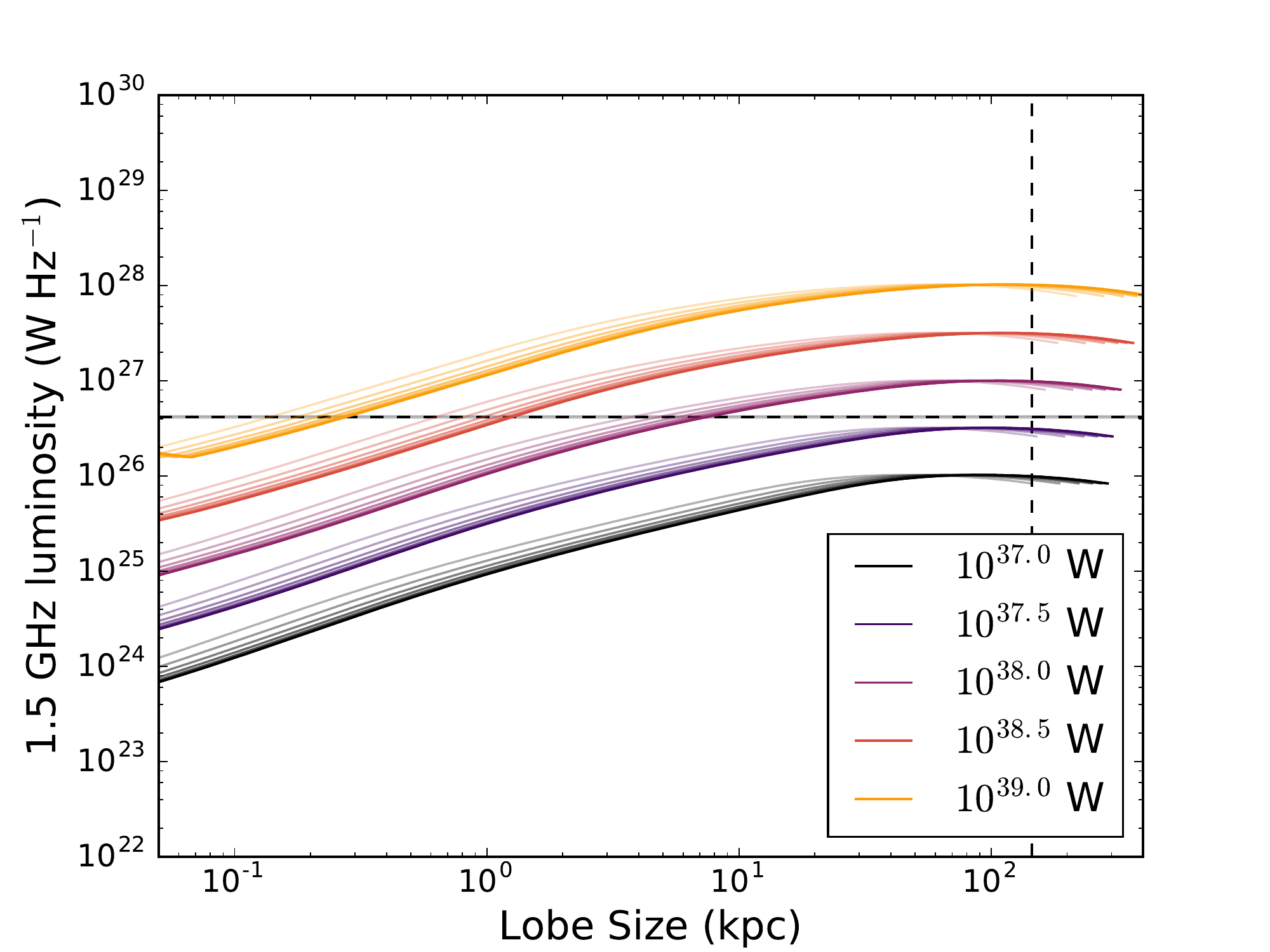}}
\caption{1.5 GHz radio luminosity evolution for a radio lobe in 3C320 (upper) and 3C444 (lower), for a range of plausible jet powers ($Q$), colour-coded as displayed on the legend, for $\zeta=0.1$. Dashed horizontal and vertical lines indicate the measured properties as given by our VLA 1.5 GHz observations. Grey shaded regions indicate 3$\sigma$ errors, based on the local RMS level on the radio map. Model lobe sizes with varying source orientation angles are given by transparent lines for each colour: solid lines represent $\theta =90^{\degree}$ (edge-on radio galaxy) and the most transparent line for each jet power representing a jet orientation of $\theta =45^{\degree}$ with respect to the observer.}
\label{figure:powersize}
\end{figure}
\begin{figure}
\centering
\subfloat{\includegraphics[scale=0.45, trim={0.35cm 0cm 1cm 1cm},clip]{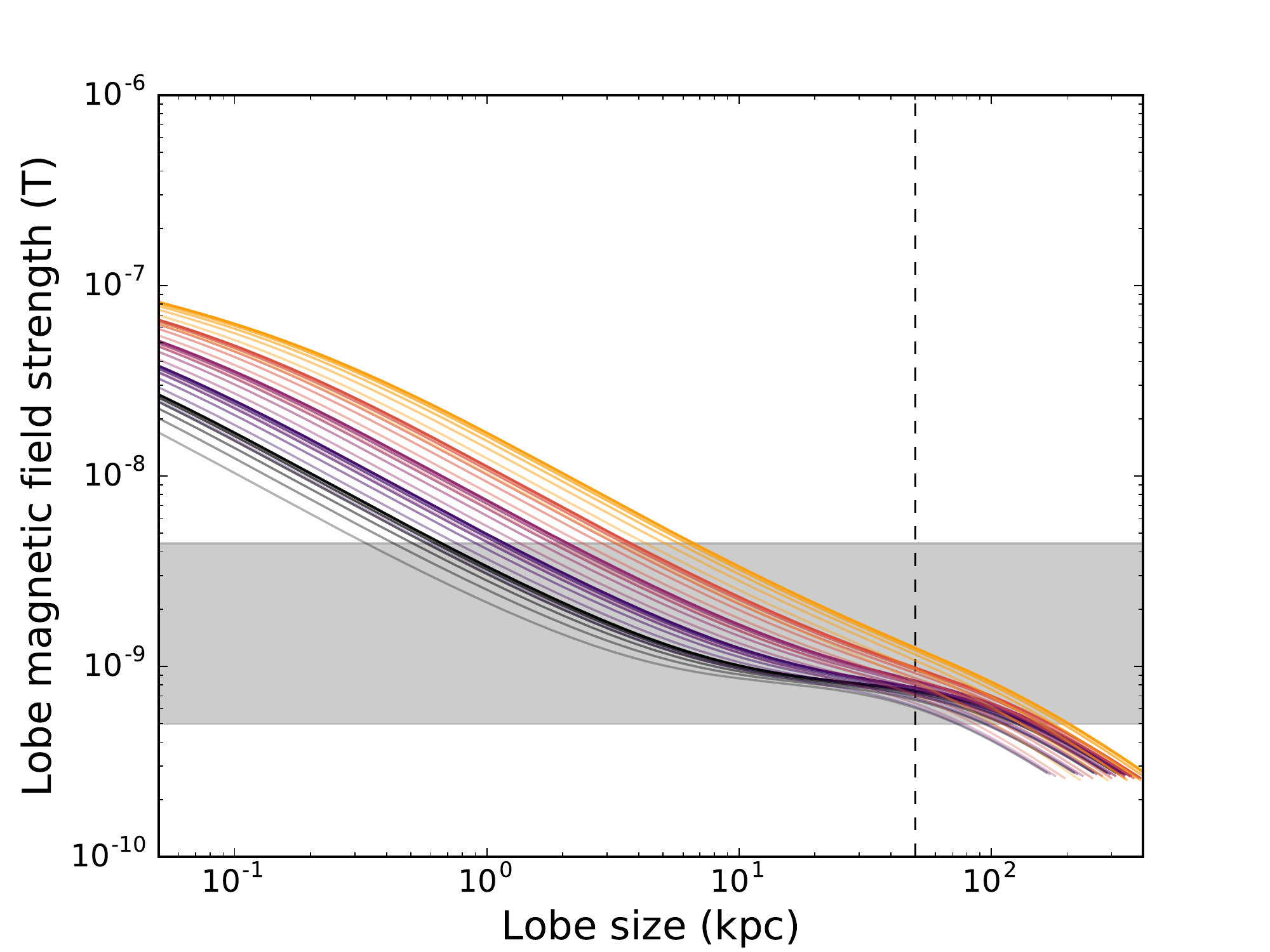}}\\
\subfloat{\includegraphics[scale=0.45, trim={0.35cm 0cm 1cm 1cm},clip]{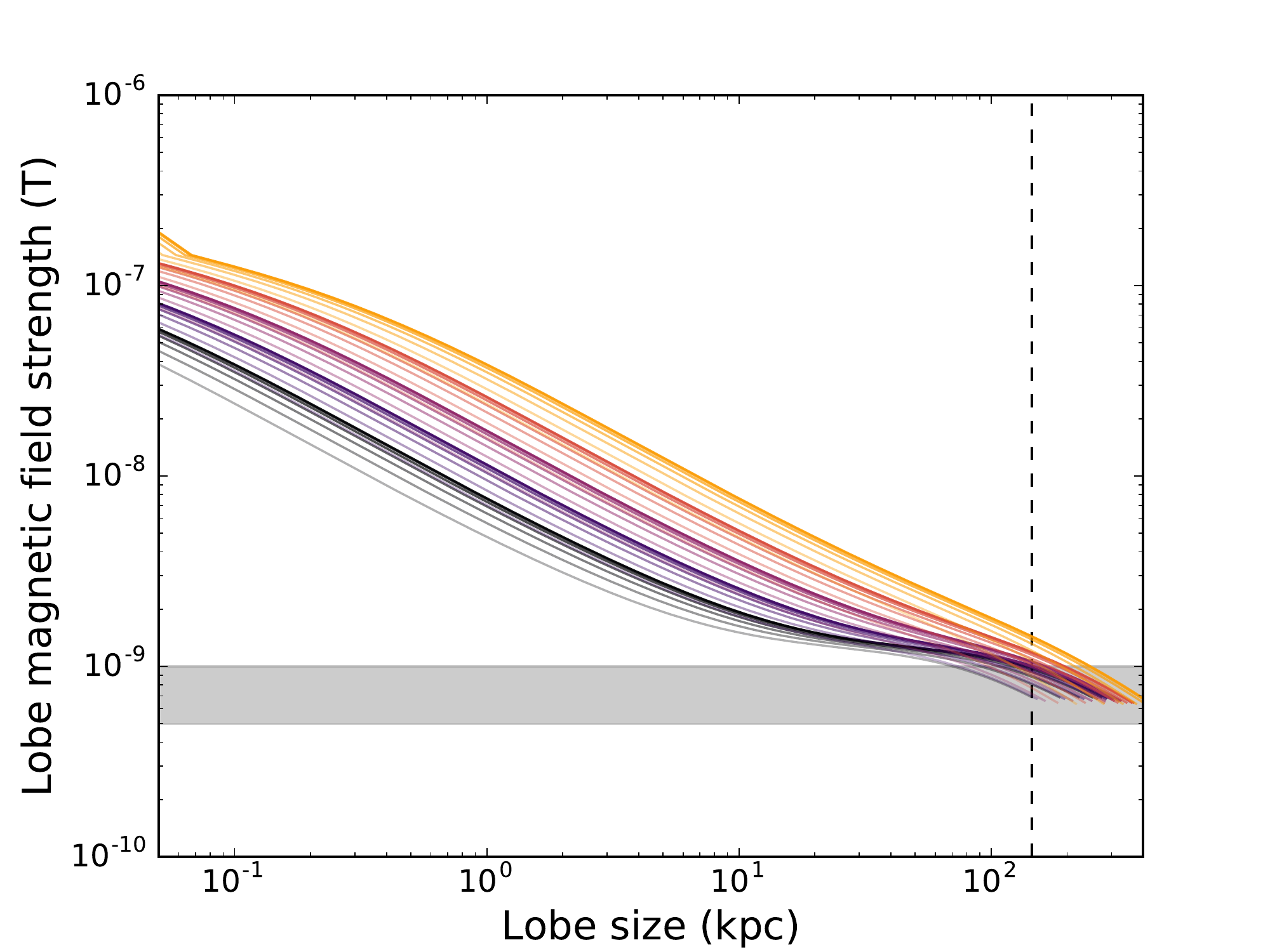}}
\caption{Magnetic field evolution for a radio lobe in 3C320 (upper) and 3C444 (lower), for a range of plausible jet powers ($Q$), colour-coded as displayed on the legend (as in Figure \ref{figure:powersize}). Grey shaded regions indicate our measured upper and lower limit magnetic fields, given by the equipartition assumption and measured inverse-Compton upper limits, respectively. The dashed vertical line and transparent coloured lines are defined as in Figure \ref{figure:powersize}.}
\label{figure:agebfield}
\end{figure}
\subsubsection{Observationally constrained shell energetics}
Example plots of the modelled luminosity-size evolution, for a range of jet powers and a range of inclination angles, at a fixed value of $\zeta$ at 0.1 for both sources, are given in Figure \ref{figure:powersize}. The general lobe luminosity evolution track is as expected from numerical simulations of powerful radio galaxies \citep{hakr13,engl16}: the luminosity starts at a minimum at very early times, then rises to a peak as the total lobe energy increases and the shell expands. The steep fall off at late times, as seen for 3C320, can be explained as the shell dynamics at this point are governed by a much lower environmental density, while radiative and adiabatic expansion losses become more prominent. We also show in Figure \ref{figure:agebfield}, using the same models as in Figure \ref{figure:powersize}, the lobe magnetic field as a function of lobe size, showing a smooth decrease of the magnetic field strength (in log space) driven by the expansion of the lobes at constant jet power.

We have over-plotted dashed lines with shaded regions showing our measured instantaneous lobe 1.5 GHz radio luminosities and physical sizes based on our VLA L-band observations (shown in Figure \ref{figure:radio_obs}). Flux densities for the luminosity calculation were obtained by drawing a region around the lobes using CASA, and 3$\sigma$ errors found from the local RMS value, propagated along to find luminosity errors (plotted as grey shaded regions in Figure \ref{figure:powersize} but too narrow to be seen in log space). Physical sizes are based on the largest angular size between the core and hotspot of the lobe showing the clearest hotspot, converted to a physical size using their redshifts. The choice of lobe does not affect the results we present here as orientation effects on the measured size are considered in our models by using a range of inclination angles $\theta$. As well as our observational constraints on the model luminosity and size, we also over-plot  our constraints on the current lobe magnetic field, given by our equipartition and inverse-Compton-based limits (shaded regions in Figure \ref{figure:agebfield}).

Figure \ref{figure:powersize} and Figure \ref{figure:agebfield} show that there must exist a number of models that may simultaneously produce the instantaneous observed luminosities within their 3$\sigma$ errors at the current size of the source within the allowed magnetic field range. Since the slope and position of the luminosity-size and magnetic field evolution is altogether degenerate between $Q_{\text{jet}}, \zeta$ and $\theta$, there will in fact be a three-dimensional distribution of the model parameters that may all produce the measured radio luminosity at the projected size within the magnetic field limits, highlighting models that are accepted by all our observational constraints simultaneously.
\begin{figure}
    \centering
    \subfloat{\includegraphics[scale=0.45,trim={0cm 0.0cm 0cm 1cm},clip]{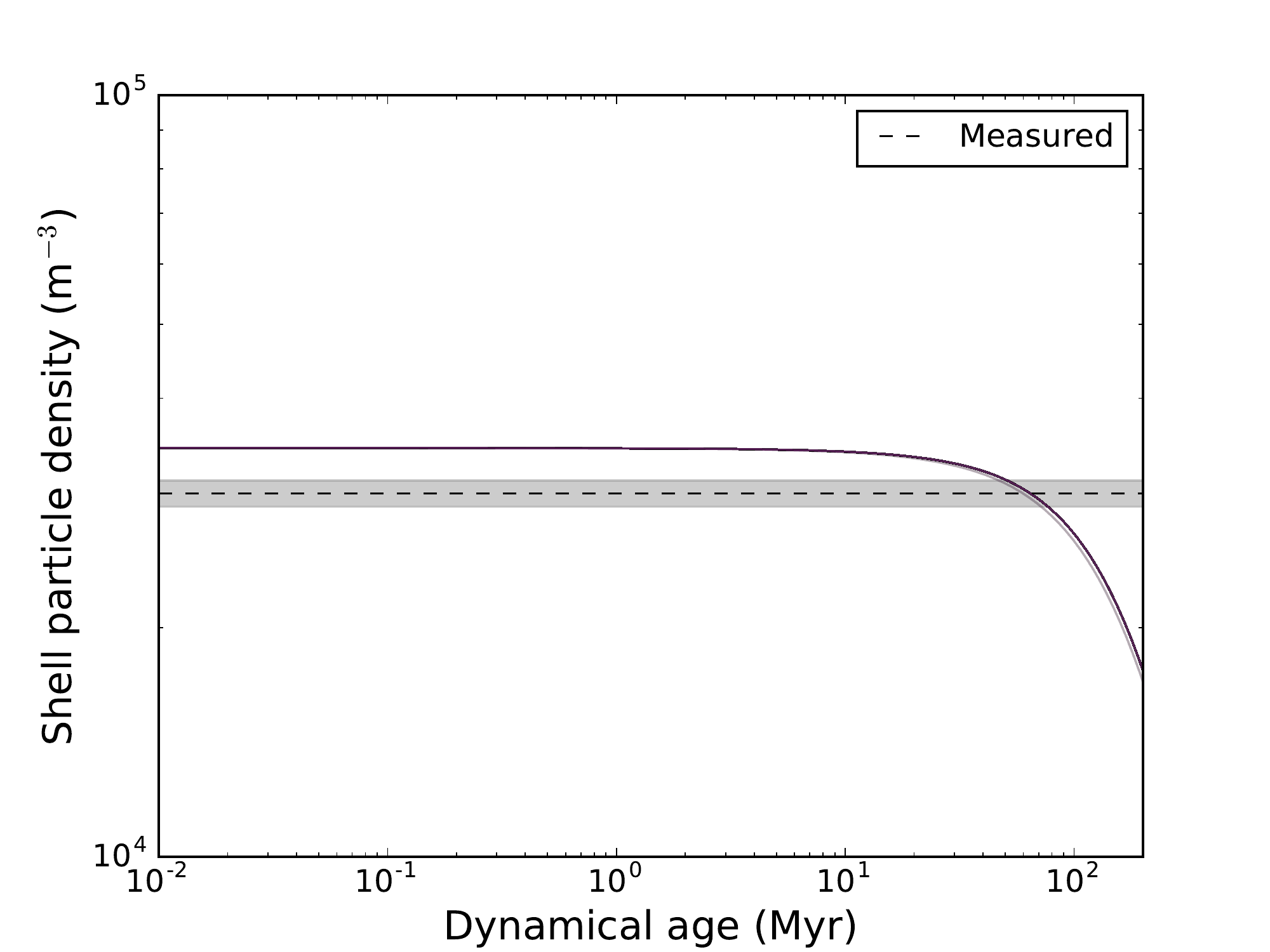}} \newline
    \subfloat{\includegraphics[scale=0.45,trim={0cm 0.0cm 0cm 1cm},clip]{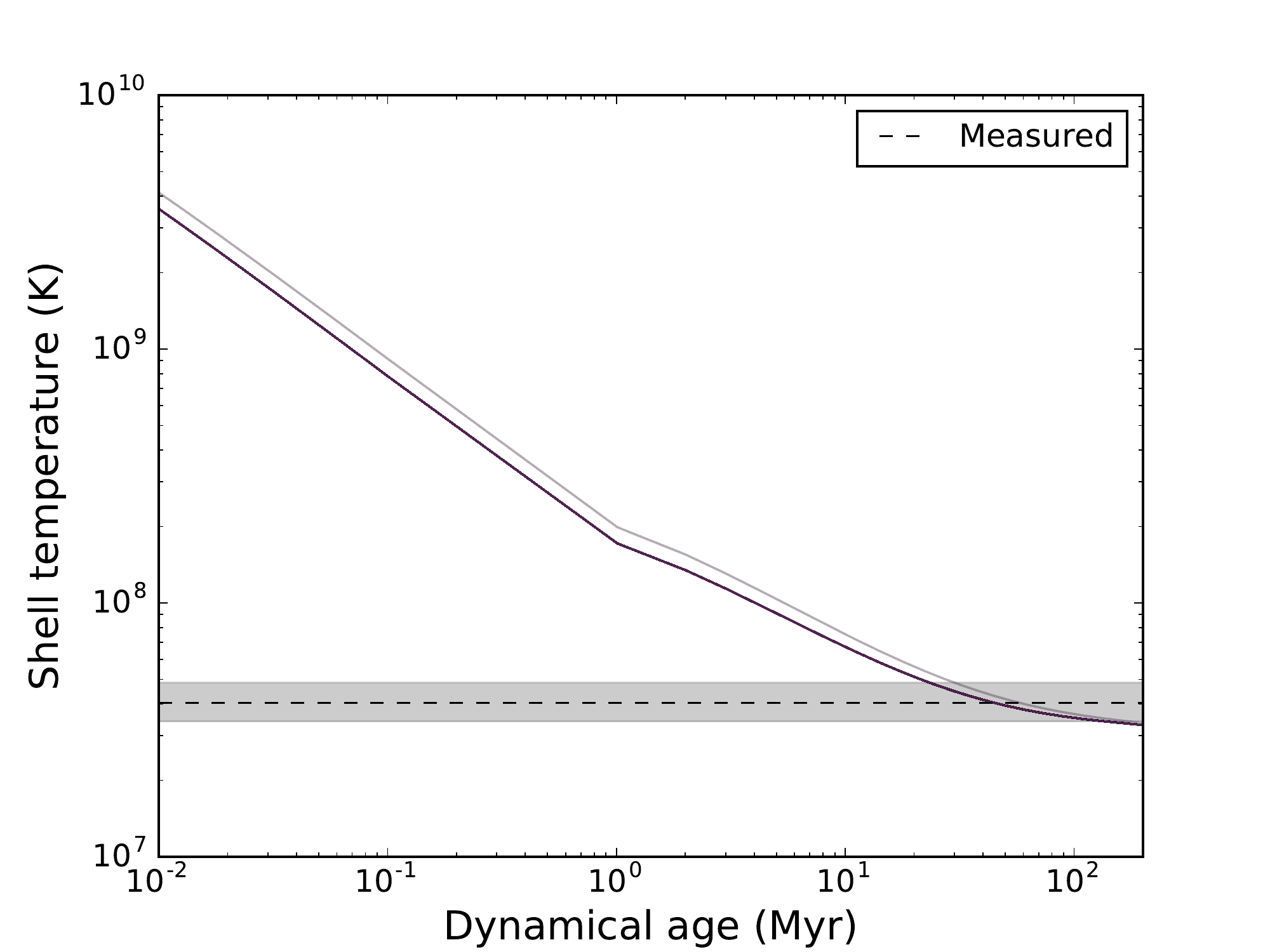}}
    \caption{Shell electron density and temperature against the source age for 3C320, for all accepted model runs (red points in Figure \ref{fig:3C320_3dmodels}). Dashed lines and grey shaded regions show our instantaneous measured values and 3$\sigma$ errors from our X-ray observations, respectively.}
    \label{fig:3C320_density_temp}
\end{figure}
\begin{figure}
    \centering
    \subfloat{\includegraphics[scale=0.45,trim={0cm 0.0cm 0cm 1cm},clip]{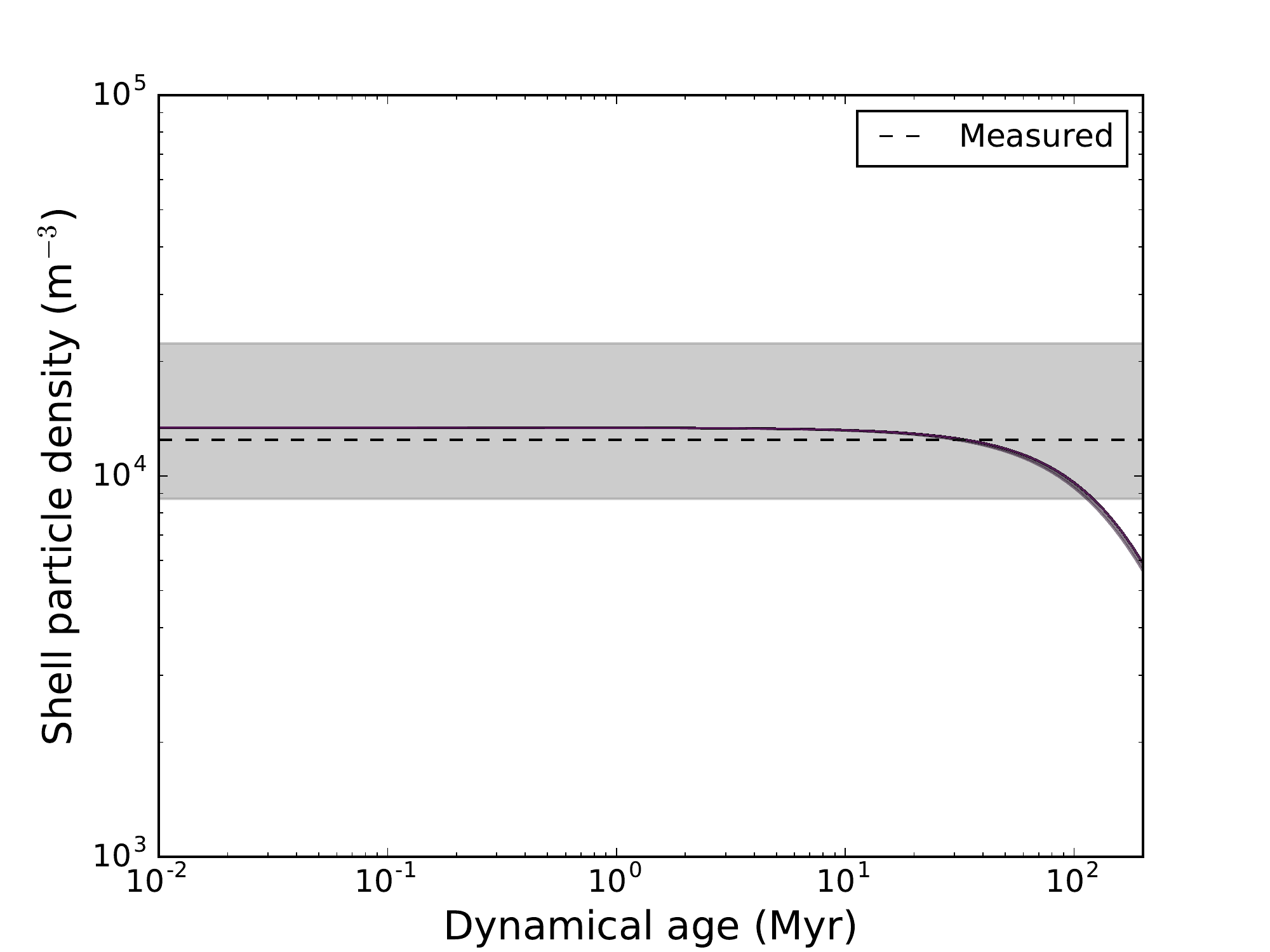}} \newline
    \subfloat{\includegraphics[scale=0.45,trim={0cm 0.0cm 0cm 1cm},clip]{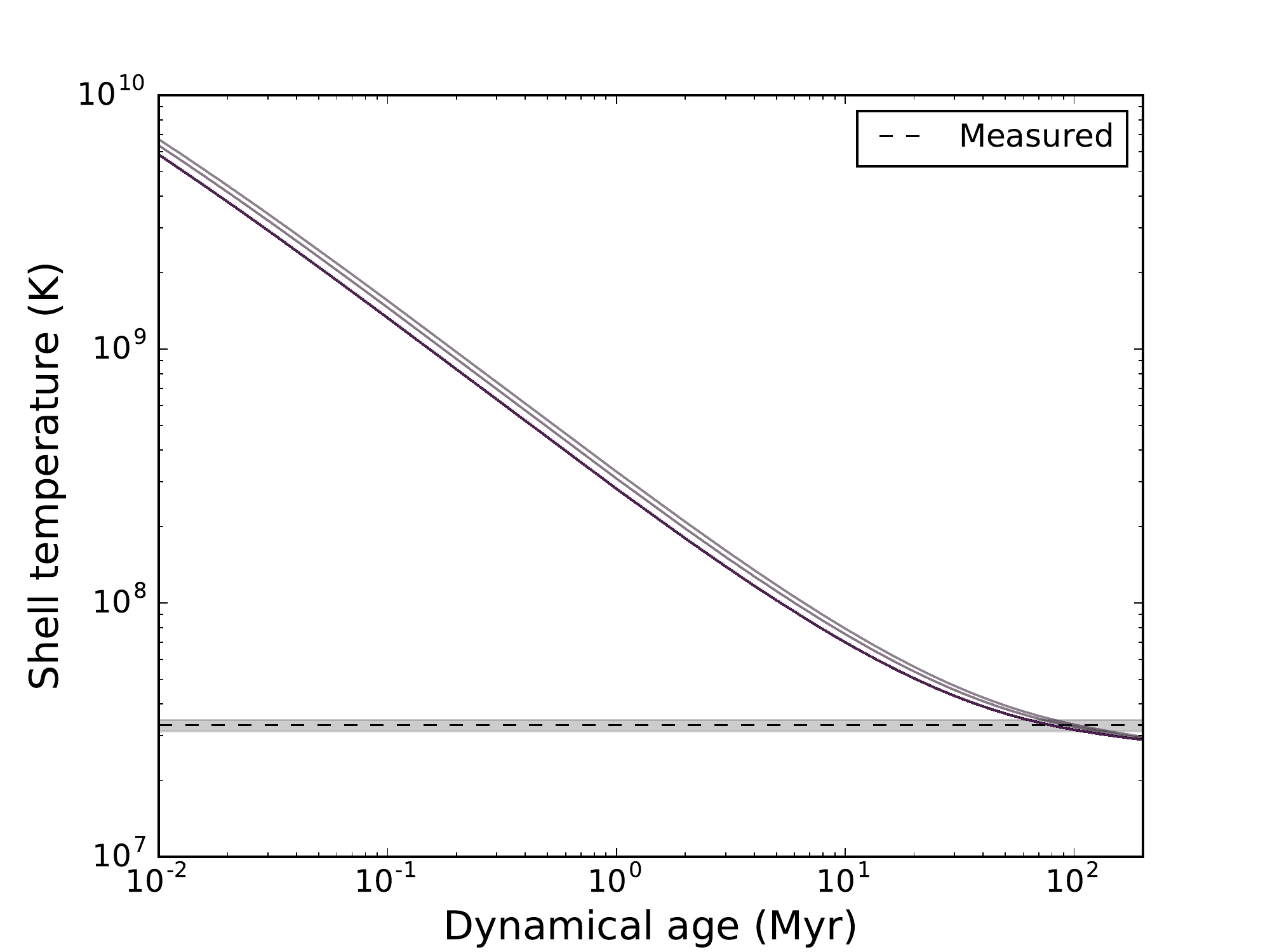}}
    \caption{Shell electron density and temperature against the source age for 3C444, for all accepted model runs (red points in Figure \ref{fig:3C444_3dmodels}). Dashed lines and grey shaded regions show our instantaneous measured values and 3$\sigma$ errors from our X-ray observations, respectively.}
    \label{fig:3C444_density_temp}
\end{figure}

\begin{figure*}
    \centering
    \includegraphics[scale=0.45,trim={8cm 4.5cm 5cm 3cm},clip]{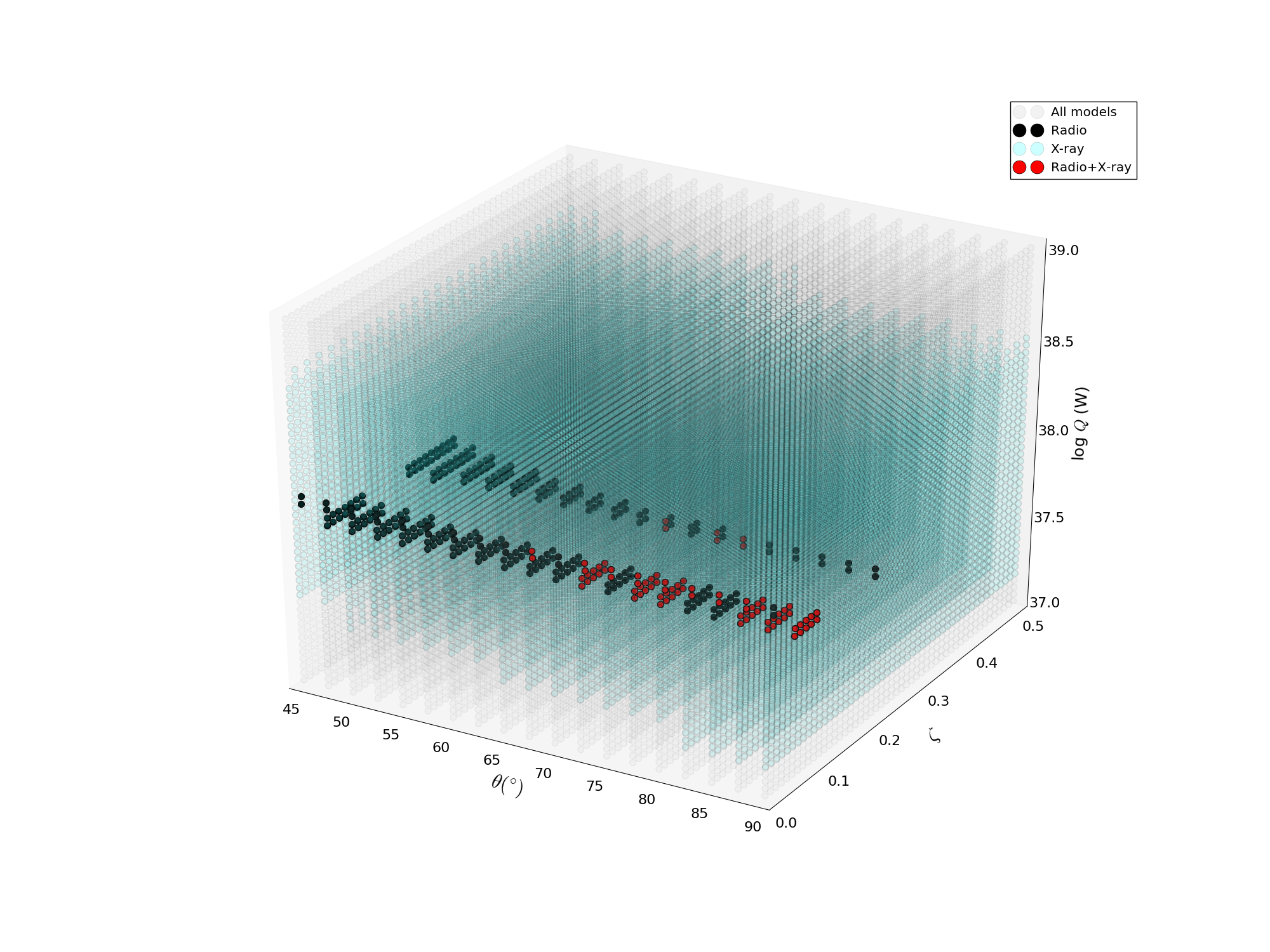}
    \caption{Three-dimensional grid displaying all model runs for 3C320 for each jet power ($Q_{\text{jet}}$), $\zeta$ and $\theta$ (grey points). Cyan points highlight the models reproducing our X-ray shock density and temperature measurements, black points highlight models reproducing the 1.5 GHz radio luminosity, lobe magnetic field and projected physical size constraints, and red models are those that match both radio and X-ray constraints.}
    \label{fig:3C320_3dmodels}
\end{figure*}

As well as the observational constraints provided by our radio data, we also apply those provided by our X-ray observations of the central shock. The analytic model also computes the thermodynamics of the shocked shell, which can be used to compare the particle densities and temperatures between those of the modelled shocked shell and those that are measured of the shock seen in our X-ray observations. The extraction of physical information from our \textit{Chandra} data for 3C320 and 3C444 is explained in Section \ref{sect:shock_measurements}. Uncertainties on the density measurements were determined by propagation of the 3$\sigma$ measurement errors from the fitted normalisation to the spectra and the uncertainties arising from projection effects on the volume calculations (see Section \ref{sect:shock_measurements}). Furthermore, for each model run, the volume uncertainties are re-calculated owing to the radio source orientation $\theta$ changing the projected lobe size that should be subtracted from the original shock volume calculation. This essentially allows the radio source orientation to vary independently to the orientations considered for the shock, with the orientations of the shock included as fixed upper and lower limits to the volume. The uncertainties, listed in Table \ref{table:xray_shock_fits}, allow us to select models that agree with our X-ray-based constraints on the shock. 

In Figure \ref{fig:3C320_density_temp} and Figure \ref{fig:3C444_density_temp}, we plot the particle density and temperature of the modelled shocked shell against age (not including the variable uncertainties on the shock volume for each model), for all models that simultaneously agree with our measured luminosity, size and magnetic field constraints. As can be seen, the shock density and temperatures vary very little between models for the range of model parameters accepted by our observations -- this is solely driven by the narrow range of jet powers constrained by our relatively small instantaneous luminosity errors, centered at $\sim 1.5\times 10^{37}$W for 3C320 and $\sim 2.5\times 10^{37}$W for 3C444. We have again overplotted dashed lines and shaded regions showing our instantaneous measured particle densities and temperatures and their errors, respectively, of the shock region from our X-ray observations (see Section \ref{sect:shock_measurements}). 
\begin{figure*}
    \centering
    \includegraphics[scale=0.45,trim={8cm 4.5cm 5cm 3cm},clip]{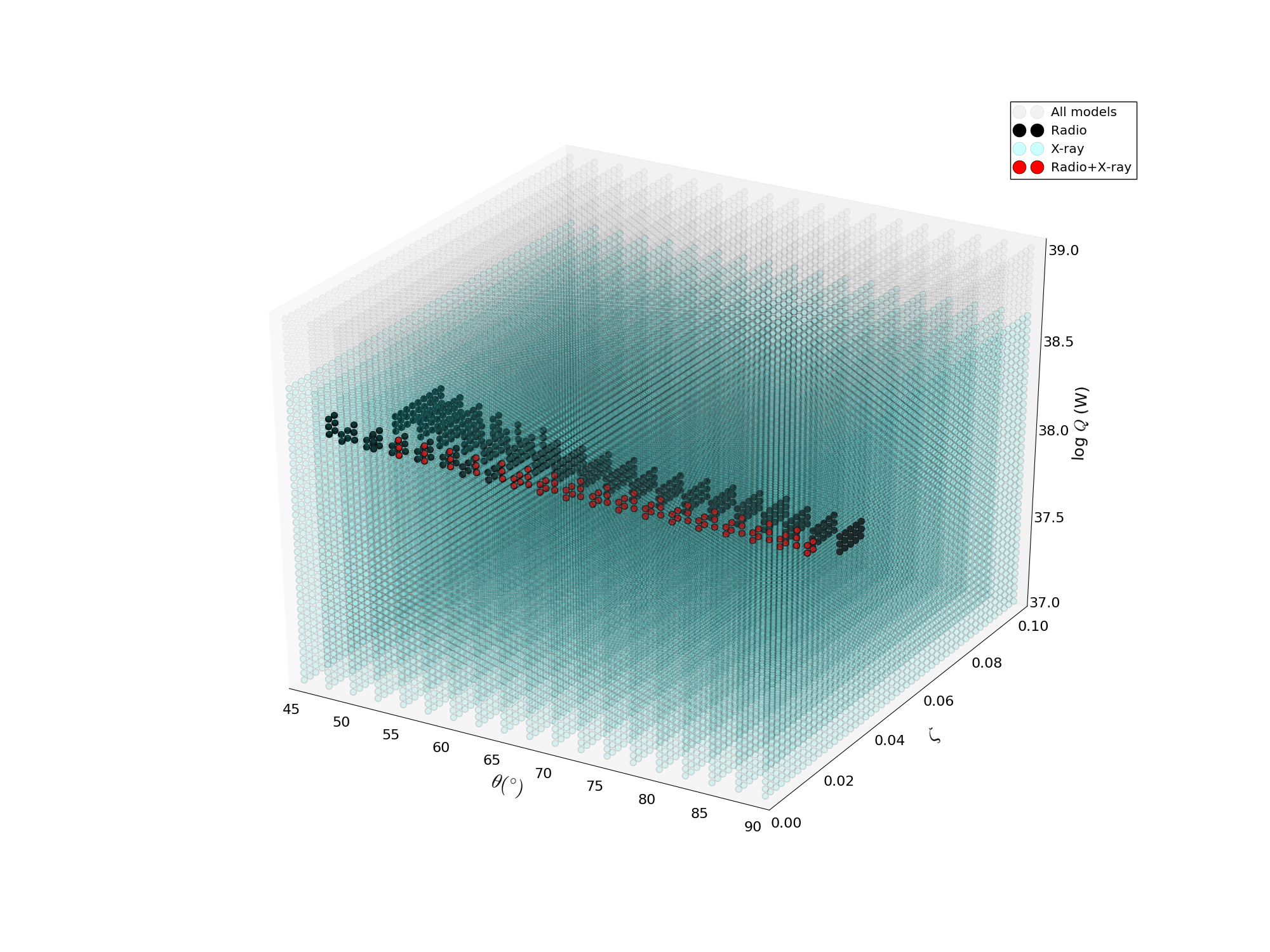}
    \caption{Three-dimensional grid displaying all model runs for 3C444 for each jet power ($Q_{\text{jet}}$), $\zeta$ and $\theta$ (grey points). Cyan points highlight the models reproducing our X-ray shock density and temperature measurements, black points highlight models reproducing the 1.5 GHz radio luminosity, lobe magnetic field and projected physical size constraints, and red models are those that match both radio and X-ray constraints.}
    \label{fig:3C444_3dmodels}
\end{figure*}
We may then use both our radio and X-ray-based observational constraints to highlight those models in the three-dimensional $Q-\zeta-\theta$ parameter space that are `accepted' by our measured instantaneous physical properties of the sources. Figure \ref{fig:3C320_3dmodels} and Figure \ref{fig:3C444_3dmodels} display these distributions for 3C320 and 3C444 respectively. Light grey points show all model runs, cyan points show models constrained by our X-ray shock observations, and black points show models constrained by our radio observations. Red points, crucially, show the models constrained by both our radio and X-ray observations simultaneously. For both sources, a similar shape in distribution of accepted models is seen, with particularly narrow ranges in jet power -- as can be seen in Figure \ref{figure:powersize} our luminosity measurement errors, by which different jet powers can be constrained, are fairly small compared to our uncertainties on other parameters (i.e $B$-field, density, temperature). For 3C320 we see a larger range in values of $\zeta$, owing to the looser constraints on the lobe magnetic field (0.5-4.4 nT) than for 3C444 (0.5-1.0 nT). Most source orientation angles $\theta$ are permitted, as expected, as we do not have fixed constraints on the true orientation angles to the plane of the sky. Interestingly, we see that the combination of radio and X-ray data for the radio source and the shock, respectively, constrain a particular set of model parameters that are more tightly constrained than the separate constraints from the data (compare red points to the black and cyan).
\subsubsection{Model dynamical and spectral ages}
\begin{figure}
    \centering
    \subfloat{\includegraphics[scale=0.45]{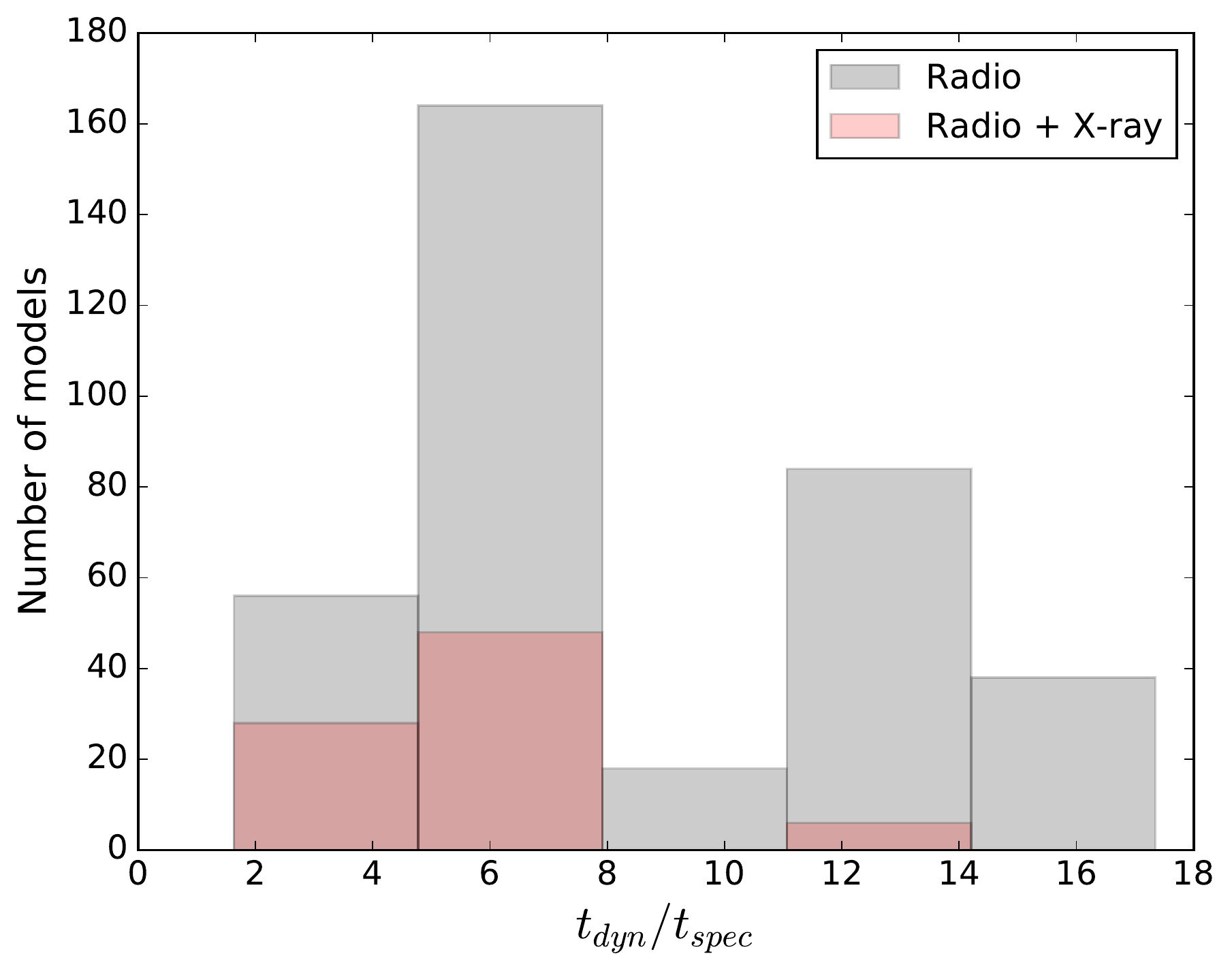}}\\
    \subfloat{\includegraphics[scale=0.45]{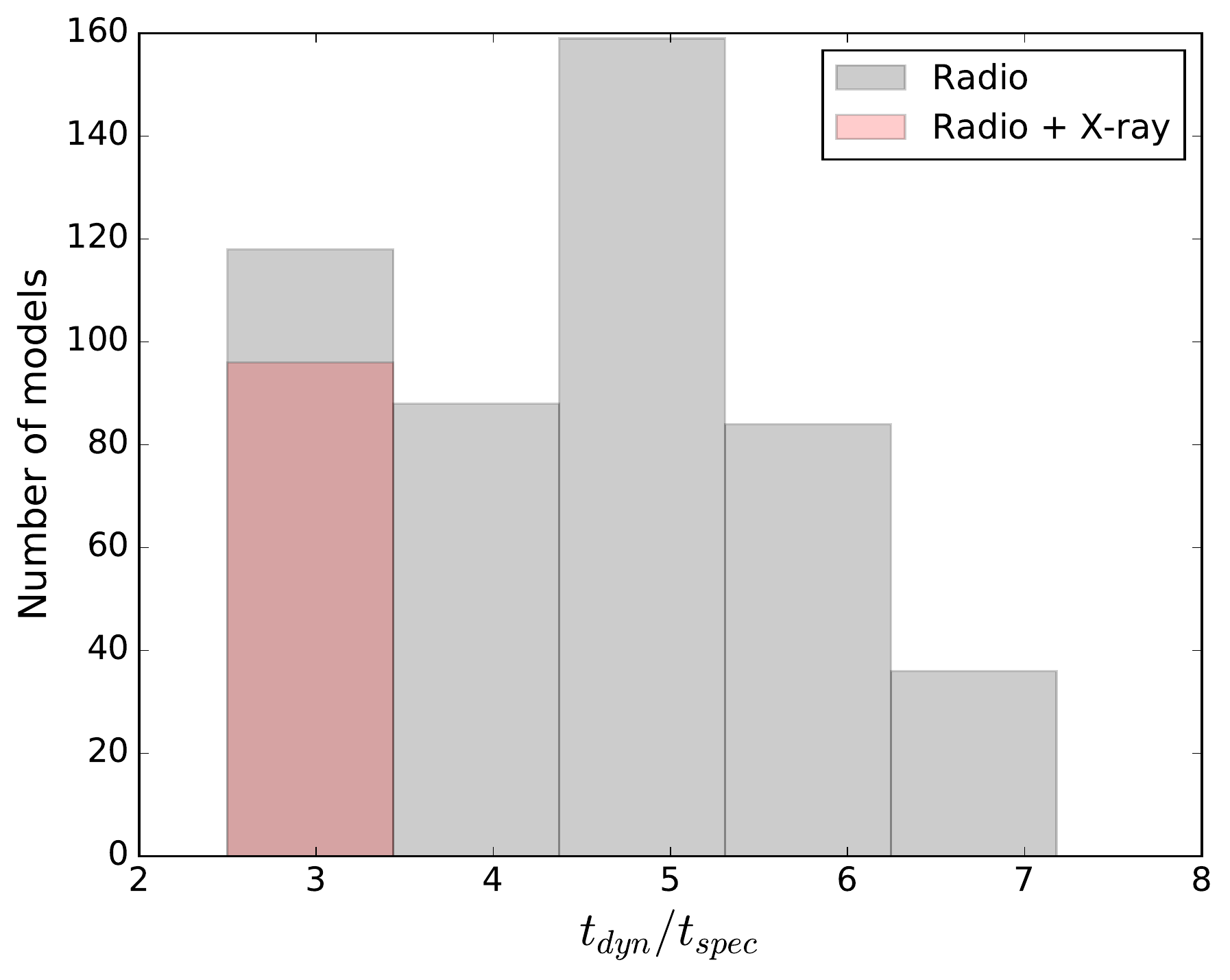}}
    \caption{Distribution of the ratio of the dynamical age to the spectral age, for our observationally-permitted models for 3C320 (top) and 3C444 (bottom).}
    \label{fig:discrepancy_distribution}
\end{figure}
For each model run, the shocked shell properties are calculated from which the lobe properties are inferred, at each time step, or, at each dynamical age between 0-200 Myr. At the time where the model properties agree with our instantaneously measured properties and their limits (red points), we compare each model dynamical age with an estimate of the spectral age of the source. As mentioned previously the crude spectral ages obtained in Section \ref{sect:specages} cannot directly be used as comparison since the true magnetic field was unknown. Instead, we use the magnetic field at the dynamical age of the source from our accepted models to re-compute spectral ages. Rather than employing {\sc brats} to re-compute spectral ages for each model magnetic field, as would be computationally and time-expensive, we simply use Equation \ref{equation:specage} for the spectral age, using the break frequency $\nu_b$ which has been obtained from the Tribble model fits in Section \ref{sect:specages}, and the magnetic field $B$ for each model. Thus, for each accepted model (red and black points in Figure \ref{fig:3C320_3dmodels} and Figure \ref{fig:3C444_3dmodels}), we obtain a model dynamical age and a spectral age based on a magnetic field associated with that model, with models being consistent with our radio and X-ray observations.

\begin{figure}
    \centering
    \subfloat{\includegraphics[scale=0.5]{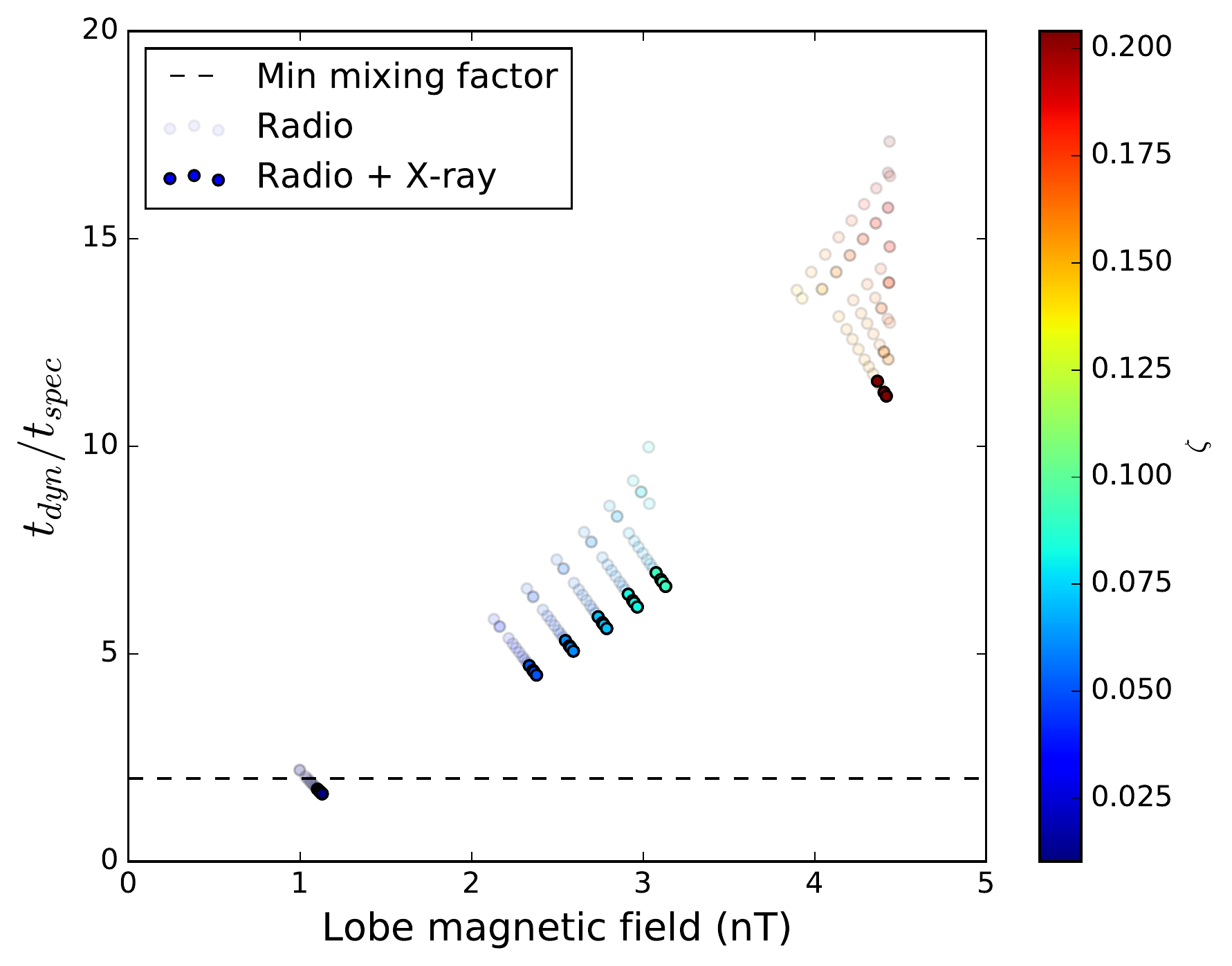}} \newline
    \subfloat{\includegraphics[scale=0.5]{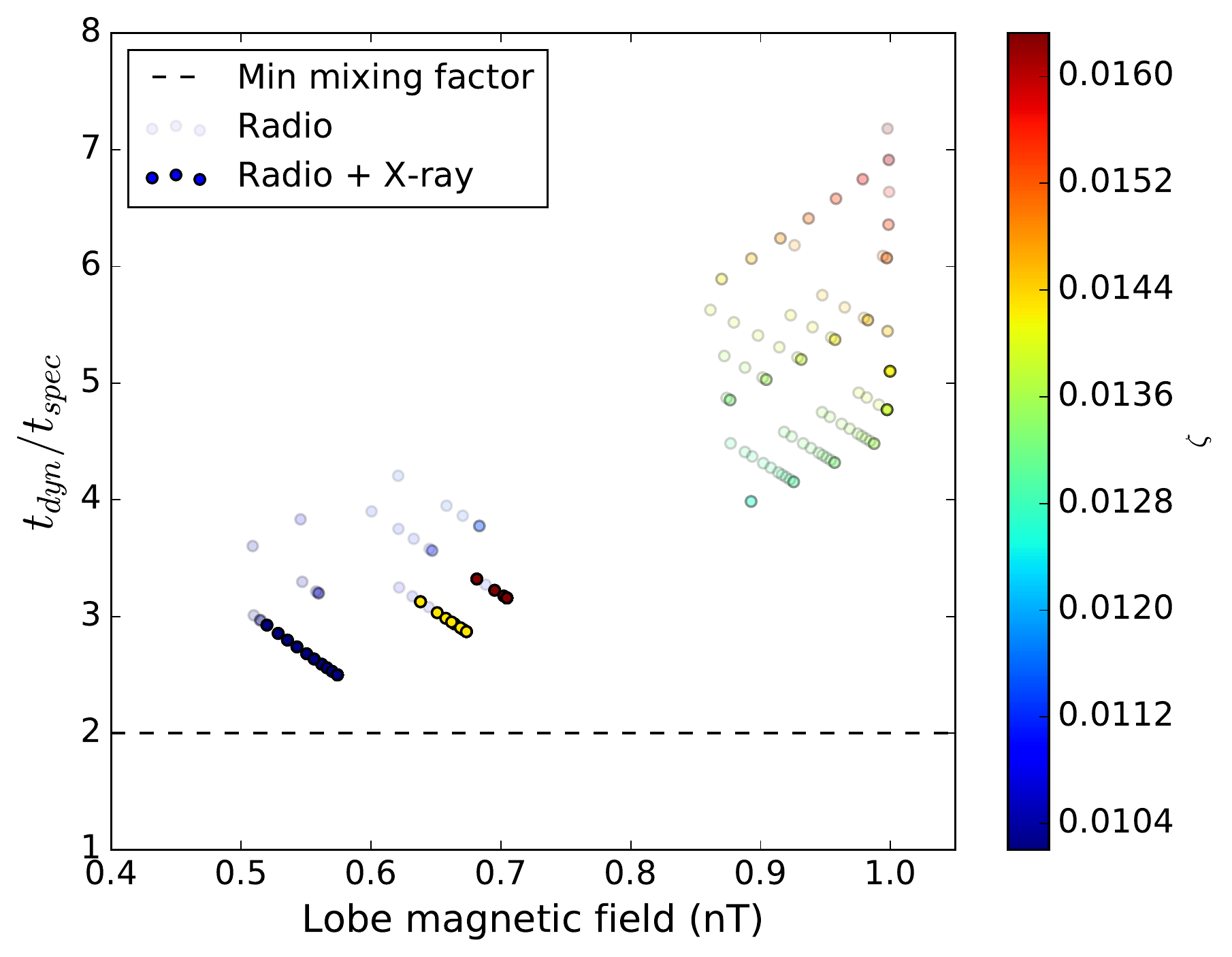}}
    \caption{Lobe magnetic field against the dynamical age/spectral age discrepancy for 3C320 (top) and 3C444 (bottom), for all accepted model runs (red points in Figures \ref{fig:3C320_3dmodels} and \ref{fig:3C444_3dmodels}). Points have been colour-coded by their value of $\zeta$. The horizontal dashed line has been drawn at an age ratio of two -- the minimum effect of electron mixing according to numerical models \citep{turn+18}.}
    \label{fig:bfield_ageratio}
\end{figure}

In Figure \ref{fig:discrepancy_distribution} we plot the distributions of the dynamical age/spectral age ratio ($t_{dyn}$/$t_{spec}$) for all accepted models (grey) and models accepted by both radio and X-ray constraints (red), for 3C320 and 3C444. Most of the models have a large age discrepancy, with a large fraction of models having $t_{dyn}$/$t_{spec}>2$. It is interesting to see that for 3C444, the models constrained by both radio and X-ray data return the best agreement between spectral and dynamical ages. This highlights the importance of obtaining accurate information on the source physical properties to obtain spectral ages, and in particular, the magnetic field strength. In Figure \ref{fig:bfield_ageratio} we plot $t_{dyn}$/$t_{spec}$ against the lobe magnetic field for the accepted models (those that satisfy our radio-based constraints in transparent colours, and those that satisfy both radio and X-ray observations in opaque colours). We colour-code each point by their value of $\zeta$, to provide a reference for each magnetic field to its value relative to equipartition ($\zeta =1$ gives an equal distribution of energy toward the magnetic field and radiating particles). It can be clearly seen that the dynamical age/spectral age discrepancy increases as $\zeta$ increases, i.e as $\zeta$ goes to 1 (equipartition) the lobe magnetic field systematically gives a spectral age that disagrees more with the dynamical age, as predicted by \cite{harw16} (hereafter \citetalias{harw16}). Spectral ages from equipartition magnetic fields are underestimated. This is consistent with inverse-Compton-based magnetic field estimates, where the population of radio-loud AGN show a departure from equipartition \citep{cros04,cros05,ines17} -- the latter study finding a median in the distribution of all sources at $B=0.4B_{eq}$. The spectral ages with the smallest discrepancy with their dynamical ages for 3C320 and 3C444 are $25$ Myr and $42$ Myr, respectively.

Our results, combined with those from other previously mentioned studies (Section \ref{sect:dynamical_ages}), suggest that the lobes of radio galaxies are not at equipartition. The total energy injected by the jets distributes more energy to the radiating particles in the lobes than the magnetic fields, than would be assumed by minimising the total energy in the lobes. As can be seen in Figure \ref{fig:bfield_ageratio}, our observations require that $B_{lobe}\leqslant 0.1B_{eq}$ for 3C320 and $B_{lobe}\leqslant 0.5B_{eq}$ for 3C444, for the best agreement between the dynamical and spectral ages. Interestingly, a lobe magnetic field that maximises the spectral age (differentiating Equation \ref{equation:specage} with respect to $B$) gives $B=B_{CMB}$/$\sqrt{3}$, indeed gives $B_{lobe} \approx 0.1B_{eq}$ for 3C320 and $B_{lobe} \approx 0.2B_{eq}$ for 3C444. \cite{shel11} find $B=0.3B_{eq}$ for the FR-II source 3C452, a similarly powerful source in a hot intragroup medium, which may suggest similar departure levels from equipartition for sources in rich environments, though we cannot state this with any significance since there are few powerful FR-IIs in cluster environments at low redshift. These and other similar departures from equipartition are discussed further in the next section.

We overplot dashed lines at a discrepancy factor of two in Figure
\ref{fig:bfield_ageratio} to highlight that the discrepancy may be due
to electron mixing -- \cite{turn+18} show, using numerical
hydrodynamical simulations of radio galaxies with radiative losses,
that the spectral and dynamical age can be discrepant by a factor of a
few, or two at minimum, due to the fact that different populations of
electrons mix in the lobes due to turbulent flows, particularly toward
the base of the lobes. Even near the core region, where the oldest
electrons exist, the observed spectrum can be dominated by younger
electrons due to mixing, increasing the observed break frequency, and
decreasing the modelled spectral age relative to the true age. The
Tribble and JP models of spectral ageing used in this study do not
take into account this physical process, as they assume that all the
electrons contributing to the model spectrum at any point all have
similar ages. The magnitude of mixing that occurs for any particular
source cannot currently be measured from observations, nor can it be
inferred for a population of sources. Even well-resolved spectral
studies are limited by this mixing factor, which can only be traced by numerical simulations. We suggest that, in the application of resolved ageing models to broad-band data, the spectral age/dynamical age discrepancy can be significantly reduced by using a correct estimate of the magnetic field strength, and a scaling factor of $\sim 2$ at minimum to account for electron mixing.

This scaling factor, however, is strongly dependent on the lobe magnetic field strength used for the spectral age calculation. The magnetic field strengths output at each dynamical age in our dynamical models are the current field strengths in the lobes at that age, as would be measured observationally as an instantaneous measurement, rather than an \textit{effective} field strength that takes into account the time-evolution of the magnetic field experienced by the same particles throughout the source lifetime. The effective magnetic field would be larger at any given time relative to the current field strength, making spectral ages younger and hence more discrepant with dynamical ages. Running the models using an effective time-integrated magnetic field, as can be output by the analytic model, we find that for a source at an age $\geqslant 1$ Myr, the effective magnetic field strength is a factor $\sim 2$ greater than the current magnetic field strength at each age. This leads to a larger discrepancy between spectral and dynamical ages by a factor $\sim2$, for each of the models shown in Figure \ref{fig:discrepancy_distribution} and Figure \ref{fig:bfield_ageratio}. The currently existing ageing models (i.e JP, Tribble) cannot determine effective field strengths for fixed electron populations while fitting for the age itself, and better models are needed to incorporate effective magnetic fields for differently aged populations. The use of the current lobe magnetic field at any dynamical age does not affect the main conclusions of this paper. 

\section{Discussion}
\label{sect:discussion}
In this section we discuss our results in probing the spectral age/dynamical age problem, and their implications on determining spectral ages of the population of radio galaxies in upcoming large-area radio surveys.
\subsection{Can we trust spectral ages?}
For 3C320 and 3C444 we have found that the spectral and dynamical age discrepancy still exists, but decreases the lower the magnetic field -- for the spectral age modelled using the Tribble model, the spectral and dynamical ages can \textit{only} be reconciled based on a relatively large departure from the equipartition magnetic field strength ($B\approx 0.1B_\text{eq}$ and $B\approx 0.5B_{\text{eq}}$) for 3C320 and 3C444, respectively. That is, a magnetic field near the inverse-Compton-based lower limits gives the best agreement for both sources (down to a factor of 2). The corresponding ratio of the lobe magnetic field to that at equipartition for 3C320 is generally lower than that estimated for other samples of FR-II radio galaxies \citep{hard98,hard02}, particularly lower than the minimum found in more recent studies \citep{cros05,ines17}. Reconciling with the work from \citetalias{harwood13}, a maximum spectral age occurs when $B=B_{CMB}$/$\sqrt{3}$, leading to a field strength of $B=0.25B_{eq}$ for 3C 436. This is also a large departure from their assumed equipartition field strengths, and would drive their spectral ages into more of an agreement to their older derived dynamical age for this source. This may suggest a common departure of $B \approx 0.1B_{eq} - 0.5B_{eq}$ for powerful radio galaxies in clusters. This is not physically unrealistic since there is no \textit{a priori} reason to believe that equipartition fields are the true field strengths in the lobes of radio galaxies, other than that they are close to the field strengths for which the total energy content in the lobes is minimized.

A major caveat in the use of observations to determine spectral ages
is that the faintest detectable synchrotron emission in a given
observation does not necessarily relate to the oldest population of
particles. It is possible that emission from significantly older
populations that have been radiating for a long period of time is
below the sensitivity limits of our VLA observations. Particularly for
3C320, as seen in Figure \ref{figure:radio_obs} or in Figure
\ref{fig:specagemaps}, the most aged emission near the core is clearly
not detected at C-band. Higher sensitivity is required in order to
capture the oldest populations which are located towards the radio
core. For 3C320, we tried including all of the highest
frequency bandwidth C-band maps (which were originally removed from
the analysis as they contained less detectable emission from the
oldest regions near the core) but this resulted in a decrease of the maximum
spectral age only by $\sim1$ Myr. While the results of this paper are
unchanged by not including the highest frequency maps, older
undetected emission might still be present. The addition of sensitive observations at even lower frequencies than those presented in this study, such as those that can be provided by the Low Frequency Array \citep[LOFAR;][]{vanh13}, will allow the detection of fainter emission that is crucial to spectral ageing studies \citepalias[e.g.][]{harw16,harw17}. 

We emphasize that these results only apply to large and powerful FR-II
sources for which lobe particle contents are well understood. The lobe energetics for FR-I sources cannot be modelled in the same way and are less well understood. Equipartition assumptions for FR-Is tend to lead to an apparent lobe under-pressure with respect to their external medium and they are therefore thought to contain a significant fraction of massive particles \citep{cros08,cros18}.
\subsection{Injection index}
For spectral ageing studies, the slope of the low-frequency power-law
spectrum was traditionally assumed to take on values around $\sim
0.5-0.7$ (e.g. \citealt{cari91}). More recent studies by
\citetalias{harwood13} and \citetalias{harwood15},
without broad-band low-frequency data, tested these assumptions using
{\sc brats} and found much steeper injection indices at $\sim
0.80-0.85$. One might expect that the addition of low-frequency data
constraining the power-law gradient would yield flatter injection
indices -- our broad-band data at 1.4 GHz for 3C320 and 3C444 give
best fit injection indices of $\sim 0.60-0.65$, closer to the
traditionally expected values. However, even with the inclusion of
MHz-frequency data, \citetalias{harw16,harw17} still find
$0.7\leqslant\alpha_{inj}\leqslant0.85$. Variations in instrument
capabilities aside (i.e broad-band vs narrow-band spectra), a range in
$\alpha_{inj}$ is routinely found in FR-II radio galaxies, but it is
interesting to discuss the physical source properties that may lead to
particular observed values of $\alpha_{inj}$, particularly since
steeper injection indices that cause younger spectral ages to be
derived may cause a discrepancy with dynamical ages.  3C320 and 3C444
are, along with Cygnus A, a group of only a handful of powerful FR-II
sources that are found in rich cluster environments. These sources are
found to have injection indices of $\alpha_{inj}\leqslant 0.65$
(\citealt[][]{carilli91} find $\alpha_{inj}\sim 0.5 $ for Cygnus A),
while interestingly, the FR-II sources in sparser environments studied
by \citetalias{harwood13,harwood15,harw16,harw17} have
$\alpha_{inj}\geqslant 0.7$. The large-scale environment may play a
role in determining the injection index we observe -- an obvious
argument for this being that FR-II sources in much
denser environments would require a higher jet power to drive through
the surrounding medium than those in sparser
  environments, thereby driving stronger Fermi acceleration at the
  terminal shock. Indeed, jet power is known to be strongly
correlated with injection index \citep{kona13} which
  supports this scenario. As mentioned, particularly with
small-sample statistics, systematic effects arising from the
measurement of injection indices must be borne in mind, and larger
samples of FR-II sources with broad-band data are required to test any
relationship between environment and injection index.
\subsection{Jet powers}
Our self-consistent dynamical models have constrained narrow ranges of
jet powers that are permitted by our radio source and X-ray ICM
observations (red points in Figure \ref{fig:3C320_3dmodels} and Figure
\ref{fig:3C444_3dmodels}). For these models we find $5.96\times
10^{37}$W$\leqslant Q_{jet,3C320}\leqslant 1.05\times 10^{38}$W and
$1.53\times 10^{38}$W$\leqslant Q_{jet,3C444}\leqslant 2.22\times
10^{38}$W. For the models with the minimum discrepancy between
spectral and dynamical ages, we have $Q_{jet,3C320}=1.05\times
10^{38}$ W and $Q_{jet,3C444} = 2.22\times 10^{38}$ W. These are
generally higher than the cavity power measurements, via a cavity
buoyancy timescale, made by \cite[][$P_{\text{cav,
      3C320}}=3.52^{+6.75}_{-1.85} \times 10^{37}$]{vags19} and
\cite[][$P_{\text{cav, 3C444}}=6.13^{+18.84}_{-2.34} \times
  10^{37}$]{vags17}, though the latter measurement has large
uncertainties and our jet power for 3C444 is close to their upper
limit. As stated in Section \ref{sect:dynamical_ages} such
instantaneous measurements are unreliable, and in this case they lead to an underestimate of the jet power, as shown by comparison to our more robust calculations. The cooling luminosity of the cluster emission surrounding 3C320 and 3C444 up to 100 kpc from the center was found by \cite{vags19} and \cite{vags17} to be $L_{\text{cool,3C320}} = 8.48^{+0.15}_{-0.28}\times 10^{37}$ W and $L_{\text{cool,3C444}} = 8.30^{+0.02}_{-0.20}\times 10^{37}$ W, respectively, suggesting that the mechanical powers delivered by the jets are sufficient to balance and overcome the radiative losses in the ICM.  

\section{Conclusions}
\label{sect:conclusions}
In order to develop the kinetic jet luminosity function (the number density of radio-loud AGN at each jet power), an accurate and reliable age is required. Given that the ages are accurately derived, and therefore the jet power, we can then essentially derive kinetic powers for all (FR-II) radio-loud AGN out to high redshifts. This will in fact become possible in the near future with the large numbers of radio galaxies that will be observed in large area radio surveys, which will have the resolution and sensitivity to produce high fidelity images of radio galaxies out to high redshift where powerful FR-IIs are more common. As a tool to determine source ages, the use of spectral ageing in the population of FR-II sources is crucial, and does not require deep X-ray observations.

In this work, with sensitive, broad-band and high resolution observations, coupled with the use of an analytic model to determine robust dynamical ages, we have probed the dynamical age/spectral age problem.
\begin{itemize}
\item We have used analytic-based dynamical models to constrain a particular set of jet powers, magnetic fields, spectral ages and dynamical ages that are self-consistent and agree with our radio observations of the radio galaxies 3C320 and 3C444, and their surrounding shocked gas as observed in X-rays.
\item The spectral ages which give the least discrepancy with respect to observationally-constrained modelled dynamical ages are $\sim 25$ Myr and $\sim 42$ Myr for 3C320 and 3C444, respectively. The minimum discrepancy factors are $\sim 2$ and $\sim 2.5$, respectively.
\item The dynamical age/spectral age discrepancy increases with the assumption of an equipartition field strength in our models. Hence, the magnetic field strength must be directly measured using inverse-Compton measurements or inferred using a suitable departure from equipartition to determine accurate age estimates.
\item Particle mixing by differently-aged populations, as is thought to occur physically due to turbulence, leads to a factor of a few discrepancy in the dynamical age/spectral age problem, as numerically calculated by \cite{turn+18}. Hence, the combination of a correct field estimate and a multiplicative factor of at least two to the spectral age will remove significant discrepancies relative to the source dynamical age.
\item When using resolved spectral ageing models, broad-bandwidth data that sample the power-law and spectral curvature due to ageing are crucial for determining accurate spectral ages. We find that the low frequency injection indices for our rich-cluster sources are flatter ($\alpha_{inj}\sim 0.6$) than those found in recent similar work for sources in sparser environments \citepalias{harwood13,harwood15,harw16,harw17}, suggesting that the environment may be linked to the observed injection index for powerful FR-II sources.
\item Jet powers based on $pdV$ estimates of cavities or instantaneous shock measurements and source ages based on buoyancy time-scales are incorrect, and lead to underestimates of the kinetic feedback of radio galaxies -- we suggest that analytic or numerical models constrained by observations are better in robustly determining the energetics and dynamics of radio galaxies. 
\end{itemize}
\section*{Acknowledgements}
We thank Elias Brinks and the anonymous referee for helpful comments in improving this paper.

This research has made use of data analysed using the University of
Hertfordshire high-performance computing facility
(\url{http://uhhpc.herts.ac.uk/}) located at the University of Hertfordshire. VHM thanks the University of Hertfordshire for a research studentship
[ST/N504105/1]. MJH acknowledges support from the UK Science
and Technology Facilities Council [ST/R000905/1].
JHC acknowledges support from the Science and Technology Facilities Council (STFC) under grants ST/R00109X/1 and ST/R000794/1. The National Radio Astronomy Observatory is a facility of the National Science Foundation operated under cooperative agreement by Associated Universities, Inc. JM acknowledges financial support from the State Agency for Research of the Spanish MCIU through the ``Center of Excellence Severo Ochoa'' award to the Instituto de Astrof\'isica de Andaluc\'ia. (SEV-2017-0709)
e-MERLIN is a National Facility operated by the University of Manchester at Jodrell Bank Observatory on behalf of STFC. The scientific results reported in this article are based to a significant degree on observations made by the Chandra X-ray Observatory, data obtained from the Chandra Data Archive, observations made by the Chandra X-ray Observatory and published previously in cited articles. This research has made use of software provided by the Chandra X-ray Center (CXC) in the application packages CIAO, ChIPS, and Sherpa. Based on observations obtained with XMM-Newton, an ESA science mission with instruments and contributions directly funded by ESA Member States and NASA









\appendix


\bsp	
\label{lastpage}




\bibliographystyle{mnras}
\bibliography{references} 




\bsp	
\label{lastpage}
\end{document}